\documentclass[10pt,conference]{IEEEtran}
\usepackage[normalem]{ulem}

\newcommand{\name}{\texttt{TopGen}}
\usepackage{cite}
\usepackage{amsmath,amssymb,amsfonts}
\usepackage{graphicx}
\usepackage{textcomp}
\usepackage{xcolor}
\usepackage[hyphens]{url}
\usepackage[braket, qm]{qcircuit}
\usepackage{graphicx}
\usepackage{caption}
\usepackage{subcaption}
\def\BibTeX{{\rm B\kern-.05em{\sc i\kern-.025em b}\kern-.08em
    T\kern-.1667em\lower.7ex\hbox{E}\kern-.125emX}}
\usepackage{color,xcolor}
\usepackage{xspace}
\usepackage{multirow}
\usepackage{booktabs}
\usepackage{lipsum}
\usepackage{soulutf8}

\usepackage{colortbl}

\usepackage{enumitem}
\usepackage{braket}

\usepackage{amsmath}
\usepackage{amsfonts}
\usepackage{url}

\usepackage{afterpage}
\usepackage{ulem}
\usepackage{setspace}

\usepackage{pifont}

\usepackage{hyperref}
\hypersetup{
    colorlinks=true,
    linkcolor=magenta,
    filecolor=magenta,      
    urlcolor=blue,
}

\usepackage{xcolor}
\usepackage{soul}
\usepackage{savesym}
\usepackage{xcolor}
\usepackage{algorithm}
\usepackage{algorithmic}
\usepackage{enumitem}
\usepackage{stfloats}
\usepackage{graphicx}
\usepackage{placeins}
\usepackage{amsmath}
\usepackage[flushleft]{threeparttable}

\newcommand{\violet}[1]{\textcolor{black}{#1}}

\begin{document}

% \title{Topology-aware Generation of Quantum Neural Network Ansatz: A Bottom-up Approach}
\title{TopGen: \underline{Top}ology-Aware Bottom-Up \underline{Gen}erator for Variational Quantum Circuits}

\author{
    \IEEEauthorblockN{
    Jinglei Cheng\IEEEauthorrefmark{2} \ \
    Hanrui Wang\IEEEauthorrefmark{4} \ \
    Zhiding Liang\IEEEauthorrefmark{3} \ \
    Yiyu Shi \IEEEauthorrefmark{3} \ \
    Song Han\IEEEauthorrefmark{4} \ \
    Xuehai Qian\IEEEauthorrefmark{2}}
    \IEEEauthorblockA{
    % \{zliang5, yshi4\}@nd.edu 
    \IEEEauthorrefmark{2}Purdue University \ 
       \IEEEauthorrefmark{4}Massachusetts Institute of Technology \
           \IEEEauthorrefmark{3}University of Notre Dame \            
    }
    % \IEEEauthorblockA{}

    % \IEEEauthorblockA{}

Corresponding authors: cheng636@purdue.edu
\vspace{-0.15in}}

\date{}
\maketitle

\thispagestyle{empty}

\begin{abstract}

% \HW{no need mention the QNN in abstract, to make the method more general to all vqa}
Variational Quantum Algorithms (VQA) are promising to demonstrate quantum advantages on near-term devices. Designing ansatz, a variational circuit with parameterized gates, is of paramount importance for VQA as it lays the foundation for parameter optimizations. Due to the large noise on Noisy-Intermediate Scale Quantum (NISQ) machines, considering circuit size and real device noise in the ansatz design process is necessary. Unfortunately, recent works on ansatz design either consider no noise impact or only treat the real device as a \textbf{black box} with no specific noise information. In this work, we propose to \textbf{open the black box} by designing specific ansatz tailored for the qubit topology on target machines. Specifically, we propose a bottom-up approach to generate topology-specific ansatz. Firstly, we generate topology-compatible sub-circuits with desirable properties such as high expressibility and entangling capability. Then, the sub-circuits are combined together to form an initial ansatz. We further propose circuits stitching to solve the sparse connectivity issue between sub-circuits, and dynamic circuit growing to improve the accuracy. The ansatz constructed with this method is highly flexible and thus we can explore a much larger design space than previous state-of-the-art method in which all ansatz candidates are strict subsets of a pre-defined large ansatz.
% The key difference compared to prior works is that 
% we try to generate the ansatz from the underlying hardware topology. 
% The effectiveness of this approach is 
% due to the flexibility of QNN ansatz.
% Because the minimized ansatz does not have specific physical meaning,
% the final states of the ansatz can be from the full Hilbert Space. 
% To make the search tractable, we propose to 
% first generate hardware compatible 
% sub-circuits with ``good'' properties, then combine the sub-circuits to form the initial ansatz. 
% We propose several optimizations to 
% compensate for the sparser qubit connection in the initially
% generated ansatz and increase accuracy.
% With this approach, the search of a huge design space is reduced to 
We use a popular VQA algorithm -- Quantum Neural Networks (QNN) for Machine Learning (ML) task as the benchmarks. Experiments on 14 ML tasks show that under the same performance, the \name-searched ansatz can reduce the circuit depth and the number of CNOT gates by up to 2 $\times$ and 4 $\times$ respectively. Experiments on three real quantum machines demonstrate on average 17\% accuracy improvements over baselines.
% finding better sub-circuits that can form the high quality ansatz.
% We evaluated our approach with 14 data sets used in 
% recent works. 
% The results show that the ansatz generated by our solution achieves decent model accuracy with ansatz that are 50\% smaller in the aspect of circuit depth. 

\end{abstract}

% The VQA algorithms contain {\em ansatz}, which 
% is a variational circuit with parameterized gates that can be iteratively updated with classical optimizers to minimize objective functions. Unfortunately, How to design good ansatz for QNN still remains a challenging open problem. Recent works adopted neural architecture search (NAS) methods to optimize the QNN ansatz. They search for ansatz architecture 
% to obtain better model accuracy. None of them considered
% quantum computer hardware topology when forming and searching 
% QNN ansatz. 

%  As a subset of VQA, Quantum Neural Network (QNN) is drawing increasing attentions because of its potential higher efficiency for Machine Learning (ML) tasks than classical counterparts.

\section{Introduction}

Quantum computing is one of the most promising emerging techniques with great future potentials~\cite{aharonov2008fault}. It exploits the exponential nature of quantum information processing, 
making it possible to solve the currently intractable problems such as 
integer factorization~\cite{shor1999polynomial}, chemistry simulations~\cite{kassal2011simulating}, quantum system simulation~\cite{zalka1998simulating,somaroo1999quantum} and large-scale database search~\cite{grover1996fast,jones1998implementation}.
Significant breakthroughs have been made over the
recent decades and different types of quantum computers have been implemented~\cite{garcia2017five,clarke2008superconducting,kane1998silicon,makhlin1999josephson}. 
However, the number of quantum bits (qubits) inside the quantum computers is still very limited. 
At the same time, the decoherence time of the qubits
and gate fidelity
seriously restrict the number of operations supported 
by the quantum machines. 
For these reasons, 
quantum computing is currently
in the so-called Noisy Intermediate-Scale Quantum (NISQ) era~\cite{preskill2018quantum}. 
The concept of {\em quantum volume} has been proposed to measure the capability of NISQ quantum computers~\cite{cross2019validating}.
Today, the most advanced quantum computers are composed of over 100 qubits~\cite{ibm_device} and the ``quantum supremacy'' has been recently demonstrated~\cite{gibney2019hello}.
% The random circuit sampling task requires the Sycamore quantum computer~\cite{arute2019quantum} to generate 1 million bitstrings with a certain quality. It is estimated that 200 seconds are needed for quantum computer, and 10 millennia are needed on the Summit supercomputer.
% The problem is easy to be solved with quantum mechanics, but is
% proven hard to simulate classically with complexity-theoretic evidence.
However, the targeting problem (random circuit sampling) is not practically useful.

A key characteristic of NISQ quantum computers 
is that the circuits are prone to decoherence, high gate errors and high measurement errors~\cite{alexeev2021quantum}. 
However, with elaborately developed quantum algorithms, we expect to see quantum supremacy in more areas such as Hamiltonian simulation~\cite{tomesh2021optimized} much sooner than other problems such as integer factorization~\cite{shor1999polynomial}. {\em Variational quantum algorithms (VQAs)} are among the most promising candidates to show quantum advantage in the NISQ era. 
Examples of VQAs include variational quantum eigensolvers (VQE)~\cite{kandala2017hardware}, quantum approximation optimization algorithms (QAOA)~\cite{farhi2014quantum} and quantum machine learning (QML)~\cite{biamonte2017quantum}. These algorithms 
typically contain {\em ansatz}, which 
is a variational circuit with parameterized gates that can be iteratively updated with classical optimizer 
to minimize the objective functions, i.e., approaching 
a certain desired state.
They can be used to solve max-cut problems, 
find the ground state energy, and perform quantum chemical simulations~\cite{lanyon2010towards}. 

Among the VQAs, the quantum neural network (QNN)~\cite{abbas2021power} is gaining more and more attention due to its potential of representing complex data. Several QML models have been proposed to exploit quantum computers in practical use cases~\cite{wu2021application,phillipson2020quantum,guan2021quantum}. In QNN models, parametric quantum circuits are used as kernels to extract features from the input data set. Liu {\em et al.}~\cite{liu2021rigorous} shows that for certain data sets with quantum nature, QML models can outperform the classical machine learning models. 

\begin{figure}[t]
\centering
\includegraphics[width=\linewidth]{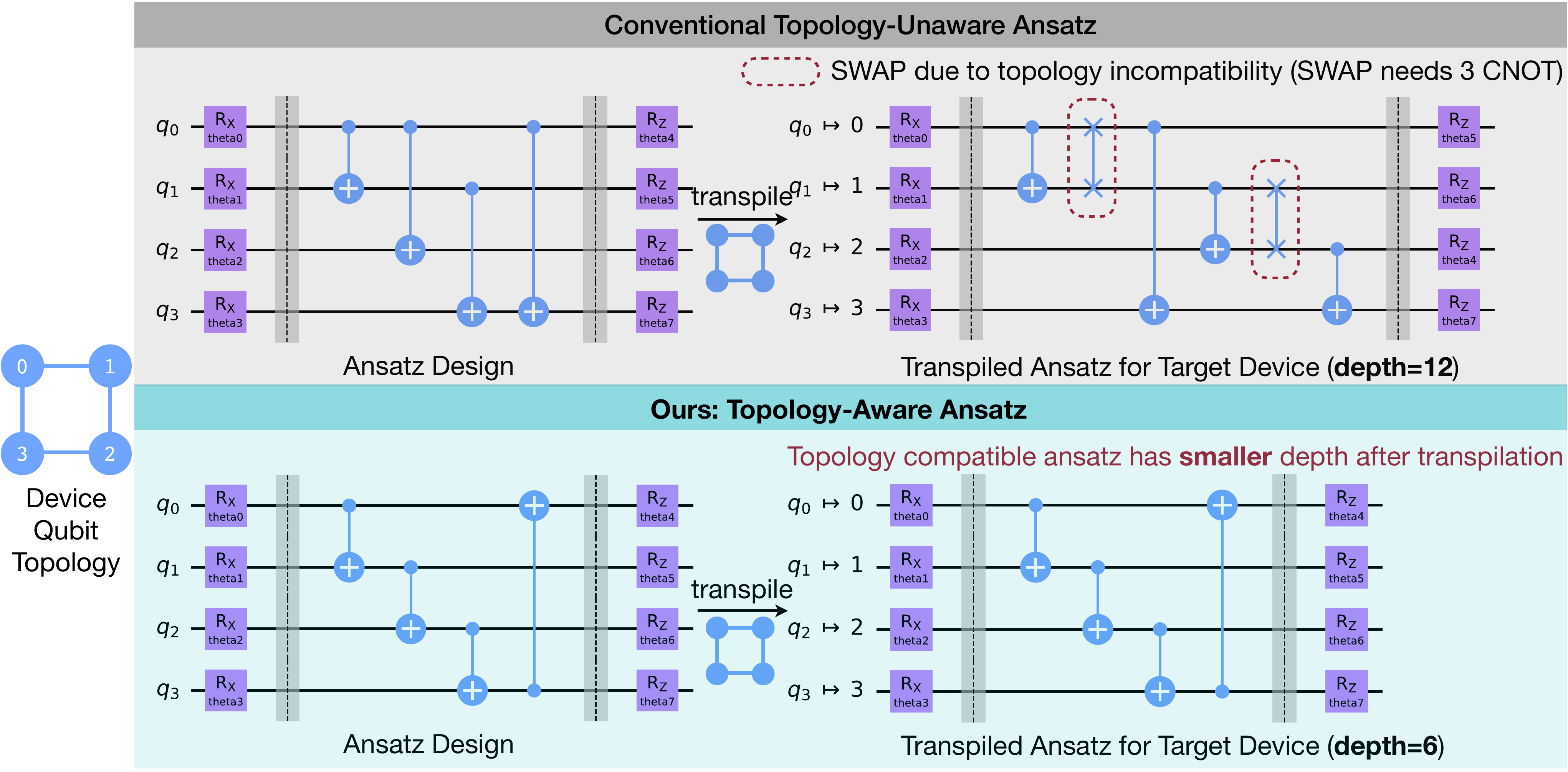}
\caption{\name \space \textit{explicitly} considers the device topology, thus the transpiled ansatz has smaller depth.}
\label{fig_teaser}
\end{figure}

% It is interesting to find the data sets with quantum nature 
% to demonstrate the advantage of QML. 
%\red{Therefore, it is crucial to find more real-world applications, especially applications with quantum nature. }

For VQAs, the ansatz is a key component:
different designs of ansatz will lead to very different performances.
The metric for the performance depends on specific applications,
it can be either energy~\cite{tilly2021variational, liang2022pan} or accuracy~\cite{wang2022chip}.
The conventional approach of choosing ansatz 
depends heavily on the applications. 
For instance, hardware efficient ansatz~\cite{kandala2017hardware} and UCCSD ansatz~\cite{bartlett2007coupled} are specially designed for VQE. 
The hardware efficient ansatz contains multiple layers of parameterized circuits. However, it has problems of trainability and barren plateaus due to its design with redundant gates.
% Single-qubit rotation gates are ins erted on each qubit and CNOT gates are applied on all possible pairs. 
UCCSD ansatz captures the essence of the electron correlations in the molecule so it is a good approximation of the ground state of a molecular Hamiltonian~\cite{grimsley2019trotterized}. 
%\red{WRITE 1-2 SENTENCES ABOUT THE INSIGHTS OF THESE DESIGNS FOR VQE}.
For QNN, designing efficient ansatz with 
low cost and high accuracy is still an open problem.

%\textcolor{blue}{
%explanation for NAS and the problems with the baseline circuits(insights on how these circutis are created)
%}

Sim {\em et al.}~\cite{sim2019expressibility} presented different ansatz derived or inspired by past studies such as hardware-efficient circuit~\cite{kandala2017hardware}, Josephson sampler circuits~\cite{geller2018sampling}, Quantum Kitchen Sinks ansatz~\cite{wilson2018quantum} and encoding circuits for QVECTOR algorithm~\cite{johnson2017qvector}.
This paper proposes a theoretical framework to characterize and compare parameterized quantum circuits based on two criteria: expressibility and entangling capability.
There are also studies on how these properties affect the performance of the ansatz~\cite{hubregtsen2021evaluation}. A strong correlation between classification accuracy and expressibility is found. A weak correlation between the entangling capability of a circuit and its classification accuracy is validated.
%\red{WHAT IS THE CONCLUSION OF THE STUDY?}

Some recent works~\cite{wang2021quantumnas,zhang2021neural} adopted neural architecture search (NAS)~\cite{pham2018efficient} in classical machine learning to optimize the ansatz for QNN. Wang {\em et al.}~\cite{wang2021quantumnas} propose to 
search the good ansatz by iteratively sampling a super-circuit. 
The evolutionary search is performed to obtain
the ansatz (sampled from super-circuit)
with the best estimated performance.
The performance estimation is conducted by using circuit simulations on noise-aware quantum simulators. Our method is fundamentally different from QuantumNAS. Firstly, QuantumNAS only considers the real machine as a black box and \textit{implicitly} considers the qubit mapping during the search so the additional SWAP gates cannot be avoided for their searched circuits. On the contrary, our method considers the real machine as white box by \textit{explicitly} designing sub-circuits with no need for SWAP insertions. Secondly, the design space of QuantumNAS is limited to subspace of a constructed SuperCircuit while ours can arbitrarily grow the ansatz.
%\red{... DESCRIBE THE RANDOM SAMPLING FROM THE SUPER CIRCUIT AND 
%USE SIMULATION TO GET THE DESIGNS WITH GOOD ACCURACY..}
Zhang {\em et al.}~\cite{zhang2021neural} accelerated the process by training an RNN neural predictor to establish the correlation between circuit architecture and circuit performance, instead 
of relying on a simulator. 
%Based on the circuit 
%Sub-circuits are sampled and simulated first. And an RNN neural predictor is used to establish the correlation between circuit architecture and circuit performance. Then a ``filled-in'' circuit is iteratively reduced and evaluated through the neural predictor to find the final ansatz architecture.}
%\red{.... DESCRIBE THIS APPROACH.}
Based on the results, the quality of the ansatz generated
by the NAS based approach is 
superior to the ones manually designed such as hardware-efficient ansatz and the ansatz with replicated layers.
%\red{FOR EXAMPLE..}
In ~\cite{wang2021roqnn} and ~\cite{liang2021can},
authors propose frameworks to mitigate the qubit noise during the learning process. The method of weights pruning has also been introduced into 
the optimization of ansatz by recent work~\cite{sim2021adaptive}.

%\violet{Experimental results show that different ansatz with proper design can reach decent accuracy at the same time. That is, there is space for ansatz optimization to include the topology information. We assume that automated design of ansatz can reach similar accuracy with less gates, which is later validated through our experiments. For other quantum programs, the space of synthesis optimization is much smaller than that of QNN programs. The flexibility of QNN ansatz allows us to consider the topology information when we are generating the ansatz for the QNN algorithm.}
\textbf{Challenge in NISQ Machines}

{\bf Problem 1: mapping overhead}.
All current attempts to optimize the ansatz are mainly focused on reducing ansatz size and improving model accuracy.
However, the topology information of underlying quantum hardware is rarely considered. 
In the compilation workflow of quantum programs such as QNN, 
the ansatz will eventually be mapped onto the physical qubits in quantum machines. Due to the sparse connection of the physical qubits, 
it is realized by inserting many SWAP gates into the original 
circuit. 
The process of compiling the circuit to match the topology of a specific quantum device is referred to as transpilation. 
We observe an average of 30\% increase of circuit depth after compilation for 4-qubit hardware-efficient ansatz. The proportion of additional SWAP gates depends heavily on the circuit structure and the hardware topology. And it will generally be increased for larger circuits due to the sparser connections between physical qubits.
%\red{For example, .. USE AN ansatz IN THE PREVIOUS WORK
%AS THE EXAMPLE, REPORT THE SIZE BEFORE AND AFTER MAPPING}
It also affects the NAS based approaches, because the best ansatz selected considering ansatz size and accuracy may not be the best when the necessary SWAP gates are inserted during mapping. 

{\bf Problem 2: search overhead}.
Generating ansatz candidates for NAS is challenging because 
the search space for such circuit is extremely large---growing exponentially with the number of gates in the circuit. 
For a complete ansatz with eight qubits and 40 gates, it is almost impossible to generate a ``good'' ansatz of such size by simply assigning random parameterized gates.
The current NAS approaches use a top-down approach, i.e.,
generating ansatz by sampling the super-circuit,
in which each ansatz candidate is 
still constrained by the super-circuit.
This approach inherently incurs the mapping overhead
because the super-circuit is oblivious to the 
hardware topology.
Generating ansatz from scratch in a bottom-up 
fashion has the potential to consider hardware
topology earlier and eliminate such overhead, 
however performing that without
the super-circuit will be more difficult. 
%Even in the NAS based approaches, sub-circuits are generated by sampling gates from a larger ``super-circuit'' . Different ways of sampling will lead to ansatz of different performance. The search space of such task is large, despite starting from a given ``super-circuit'' . Obviously it will be more difficult to create efficient ansatz from scratch.}
%\red{NEED TO DESCRIBE SOME CONCRETE INSIGHTS ON WHAT IS DONE
%IN THE SEARCH, WHY THERE ARE SO MANY POSSIBILITIES. YOU CAN WRITE
%THE FEW SENTENCES USING THE SPECIFIC NAS APPROACH BASED ON SAMPLING FROM
%SUPER CIRCUIT. }

\textbf{Drawbacks of state-of-the-art approach}
\violet{
Recent works~\cite{wang2021quantumnas, grimsley2019adaptive, tang2021qubit, van2022scaling, grimsley2022adapt} have proposed different methods to tackle the challenges stated above. 
However, they have different drawbacks and limitations respectively.
Adapt-VQE~\cite{grimsley2019adaptive} has demonstrated the effectiveness of a ``growing'' algorithm in VQE task. The appended circuits originate from the Pauli operator with the largest gradient during the training process.
The growing algorithm designed in adapt-VQE requires prior knowledge about the desired quantum states, which corresponds to the molecule's lowest energy state. Namely, there exists a guided way to generate a good ansatz for VQE, as demonstrated by the golden standard UCCSD~\cite{grimsley2019trotterized}. 
% However, for machine learning tasks that require QNN to solve, such as image classification, the training consumption required by the framework itself is already large, and we cannot have a large cost for the growing algorithm. 
Machine learning tasks such as image classification are addressing more complicated problems, where we have nothing similar to Pauli operators from VQE algorithm. 
Namely, we have no prior knowledge on what the optimal ansatz would be. In this case, adapt-VQE cannot be adopted to find the optimal ansatz. 
Therefore, we need to search for better ansatz architectures.
Besides, adapt-VQE evaluates its ideas with numerical simulations, whereas we test our methods on real-world quantum devices and validate the improvements shown in simulation results. 
% And adapt-VQE does not consider the realistic environment of real quantum machines, we can always test the feasibility of a algorithm on simulator, but we cannot demonstrate the quantum supremacy on a quantum simulator which simulate by classical computer. 
% There will be some major gaps and challenge to actually address on the real NISQ machine.
% QuantumNAS \cite{} states a supercircuit training property and demonstrate the advantage of the proposed algorithm on both VQE and QNN,
QuantumNAS~\cite{wang2021quantumnas} provides a top-down method to create the ansatz circuits. The ansatz circuits are extracted from super-circuits with iterative evaluations, where extra computation overhead is introduced. 
% however, we have noticed QuantumNAS required a large number of quantum gates, whereas, we have limited decoherence time in NISQ machine. 
To mitigate the drawbacks and limitations of these approaches, we provide our solution \name \space to find better ansatz for QNN.}

{\bf Our solution: a bottom-up approach based on sub-circuits}:
In this paper, we propose \name \space, a novel approach that addresses both 
problems at the same time. 
Instead of performing NAS from scratch at the gate level, 
we start from the {\em sub-circuits} as the building blocks
to narrow the search space, making the optimization problem
more tractable. 
The sub-circuits are aware of the qubit topology as shown in Figure~\ref{fig_teaser}.
This approach can avoid the mapping overhead by 
ensuring quantum hardware-compatibility of the sub-circuits
{\em by design}, i.e., there is no need to insert SWAP gates
to the best ansatz generated. 
It can ensure good accuracy by 
generating sub-circuits with ``good'' properties, which 
can be measured by the expressibility and entangling capability
as in ~\cite{hubregtsen2021evaluation}.
With this approach, the search cost is the cost
of generating the set of ``good'' sub-circuits.
The size of the sub-circuits
can naturally explore the trade-off between
the quality and cost of the search. 
If the sub-circuits are too small, they 
cannot fully explore the Hilbert Space. 
With limited search space, we may fail to generate and find the 
ansatz that is good enough. 
On the other side, if the sub-circuits are too large,
we will suffer the similar challenge of the giant search 
space as for the current NAS based approaches. 

The effectiveness of this approach comes from the flexibility of QNN ansatz. Because the optimal ansatz does not have physical meaning,
the final states of the ansatz can be from the full Hilbert Space. In comparison, the space of circuit synthesis 
for other quantum programs is much smaller than that of QNN programs because the physical meaning is reflected in the ansatz design. For example, the ideal ansatz for VQE is the approximation of the lowest energy states.
%\red{EXPLAIN REASON}.
The recent experimental results~\cite{wang2021quantumnas,wang2021roqnn,zhang2021neural} confirmed
such flexibility, showing that different properly designed 
ansatz can all reach decent accuracy.
The flexibility and large search space
indicate the potential of considering quantum
hardware topology information in ansatz design.
Our results show that the ansatz generated with our sub-circuit
based approach can reach better accuracy with less number of gates and 
latency.

 %Therefore, we decide to start from hardware-compatible sub-circuits. In this way, the design space is narrowed and the optimization problem becomes tractable. During the process of generating hardware-compatible sub-circuits, we need to pay extra attention to the size of the sub-circuits. If the sub-circuits are too small, then the sub-circuits cannot fully explore the Hilbert Space. That is, the design space is so limited, we will probably fail to find the solution that reach the expectation. At the same time, the sub-circuit cannot be too large. Since this will violate our ideas of narrowing the design space. With large sub-circuits, we will encounter a giant design space. 
 
% We need more samples to obtain sub-circuits with ``good'' properties. More sub-circuits of larger size will lead to heavy computation overhead, which is undesired. Therefore, we make sure that the sub-circuits are of proper size when we generate them based on hardware topology.

%In contrast, our design emphasizes the importance of the underlying topology information, and the ansatz is born to be hardware-compatible. 

%\textcolor{cyan}{
%Why generating a complete ansatz from scratch is not plausible, where does the flexibility of QNN ansatz come from?
%}

{\bf Contribution}.
The goal of this paper is to find the
better architecture for QNN ansatz and demonstrate in both simulator and real NISQ machine.
To realize the sub-circuit based bottom-up approach, 
we propose the following steps. 

\begin{itemize}[leftmargin=*]
    \item {\bf Sub-circuit generation.} 
    We generate a group of sub-circuits that are compatible with hardware topology. Different gates are randomly selected and inserted to form the sub-circuits. The quality of sub-circuits are measured by
    the expressibility and entangling capability. 
    %The principles to compute these criteria are discussed in details in Section \ref{eval}. The algorithm is described in Algorithm \ref{sub_gen}.
    
    \item {\bf Ansatz construction. } 
    After a library of sub-circuits is generated, 
    the sub-circuits are used as building blocks to construct ansatz. 
    The ones with superior performance will be selected and combined to generate an initial version of the QNN ansatz.

    \item {\bf Optimizations. } 
    It is likely that the initial combined ansatz does not reach the required accuracy. We propose several techniques to
    recover the accuracy loss by extending the ansatz: 
    ``stitching'' the sub-circuits to enhance the entanglement;
    ``growing'' the ansatz to improve accuracy; and
    pruning the gates with small parameters to reduce ansatz size
    without affecting accuracy. 
    
\end{itemize}

{\bf Evaluation highlights. }
We evaluated our approach with 14 data sets used in recent works. 
The results demonstrate that sub-circuits of better properties can help improve the performance of the combined ansatz. The ansatz generated by
our approach achieves an average of 13\% and 7\% accuracy improvement over random ansatz and manually designed ansatz, respectively. The sizes of the generated ansatz are similar to baselines before compilation. 
But thanks to the hardware compatibility, after compilation, 
we observe around 50\% reduction in circuit depth and up to 75\% reduction in the number of CNOT gates.

%The rest of the paper is organized as follow: Section \ref{back} gives the basic background of quantum computing and states our motivation; Section \ref{over} presents an overview of our bottom-up approach; Section \ref{bua} describes how sub-circuits are created, evaluated and assembled into topology-aware ansatz; Section \ref{opt} describes possible optimizations for the topology-aware ansatz; Section \ref{eva} gives results on our ansatz, previous work is related in Section \ref{rel} and we conclude our work in \ref{con}.

\section{Background}
\label{back}

\subsection{Quantum Compilation Workflow}

\begin{figure}[t]
\centering
\includegraphics[width=\linewidth]{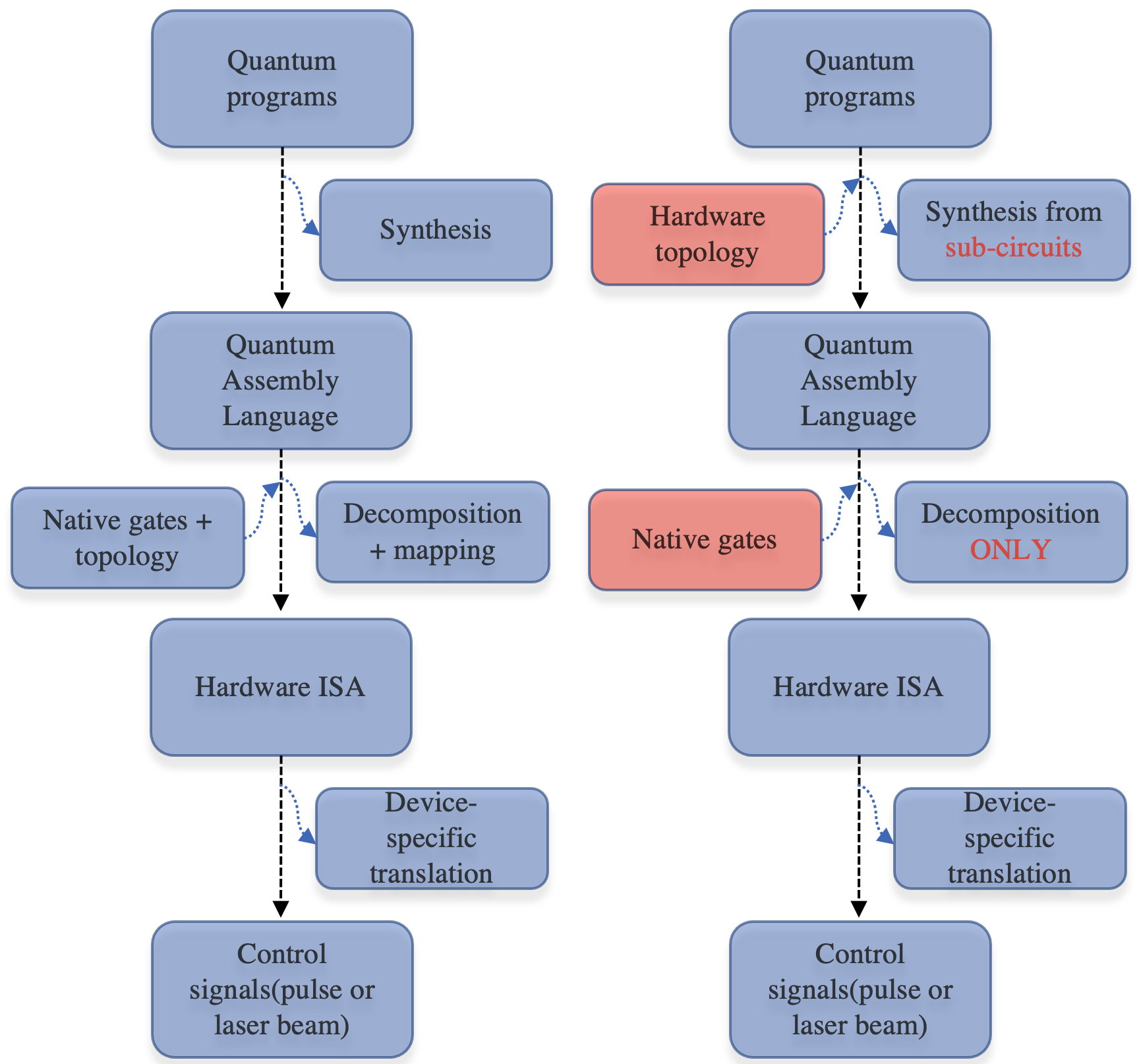}
\caption{Current compilation workflow for quantum programs compared with the proposed approach. 
% Quantum programs are first synthesized into quantum circuits. Then SWAP gates are inserted into the circuits to make it compatible with the topology of physical qubits. This process is referred to as ``mapping'' . After the mapping process, the circuits are decomposed into native gates that are supported by the quantum hardware. In the final stage, the circuits are packed and sent to quantum computers for execution. 
In \name, the hardware topology is considered during synthesis. Therefore, the ansatz needs no extra SWAP gates after the program is synthesized.}
\label{fig_compilation}
\end{figure}

Quantum programs need to be compiled before being executed on 
quantum hardware. The programs are first translated into the quantum assembly language such as OpenQASM~\cite{cross2017open}. After the program is composed into only single-qubit and two-qubit gates, the mapping algorithm will insert SWAP gates into the quantum circuits to ensure that the program is compatible to the hardware topology
of a specific quantum computer. Since the basic gates used in OpenQASM are usually different from the native gates supported by quantum computers. The quantum circuits need to be further decomposed into native gates. In different quantum computer systems, the final transitions from native gates to control signals are different. For IBM quantum computers, the control signals for superconducting qubits are microwave pulses. IBM provides Qiskit~\cite{qiskit} as a software development kit (SDK),
in which the transpilers inside Qiskit correspond with the compilation workflow, including mapping, gate decomposition, etc. 
% After the compilation is completed, instructions will be packed as qobj and sent to IBM's quantum device. 
% The results will be returned after the execution of program is accomplished. 
The traditional workflow is shown on the left of Figure~\ref{fig_compilation}.

% \begin{figure*}[h!tb]
% \vspace{-7mm}
% \centering
% \includegraphics[width=0.9\linewidth]{Figure/baseline_0_0.png}
% \caption{The baseline ansatz are selected from\cite{sim2019expressibility} with the best ``performance''. These circuits are derived or inspired from ansatz designs of other works targeting different tasks~\cite{wilson2018quantum,wilson2019quantum,geller2018sampling,2020,johnson2017qvector}.}
% \label{fig_base}
% \end{figure*}
\begin{figure}[t]
\centering
\includegraphics[width=\linewidth]{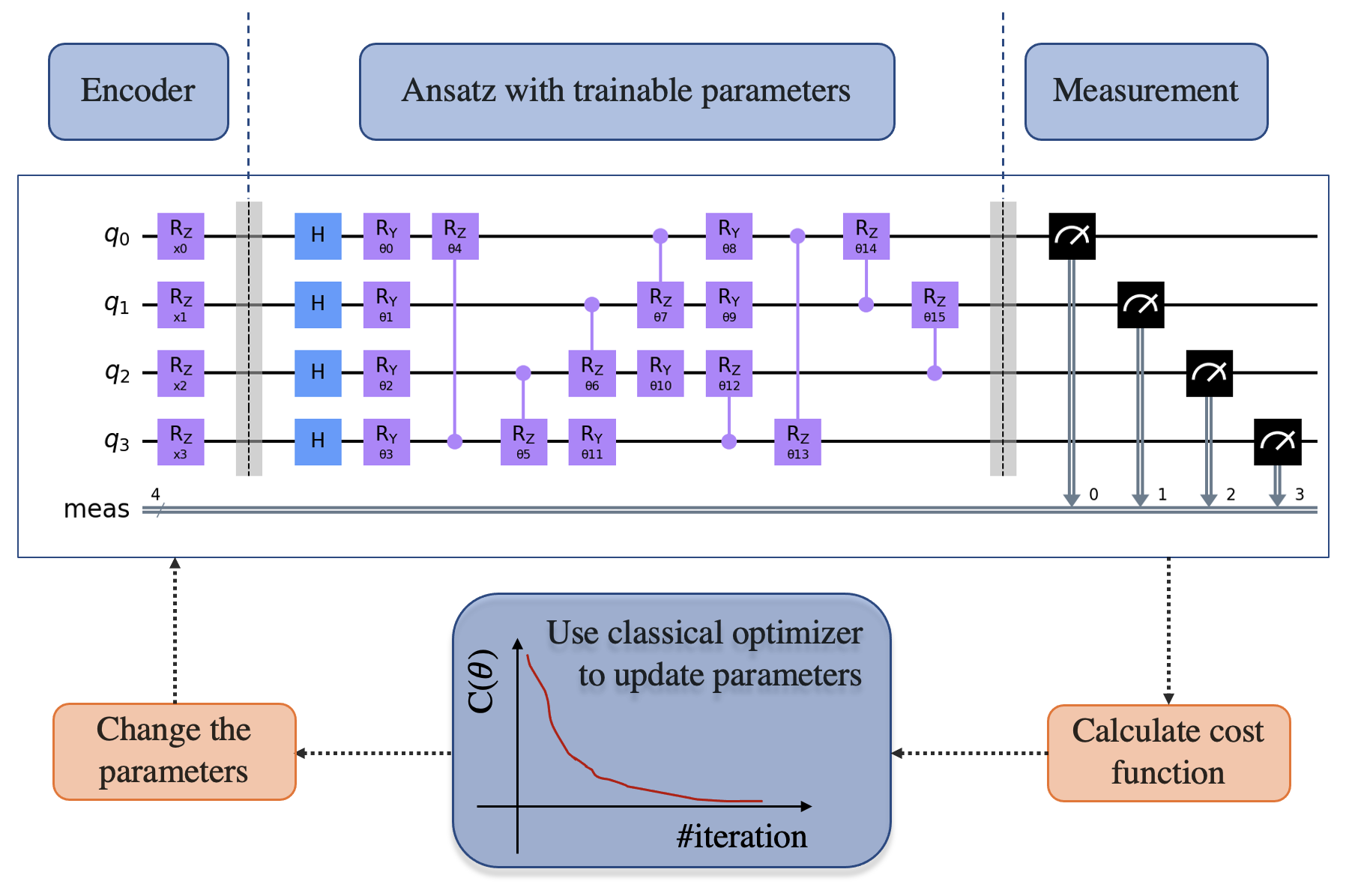}
\caption{QNN programs are composed of several parts: data encoder, ansatz, measurement and classical optimizer. In each iteration, the gradients of the parameters are calculated and the parameters are updated.}
% \vspace{-1mm}
\label{fig_qnn}
\end{figure}

\subsection{Variational Quantum Algorithms}

In the NISQ era, VQAs are among the most promising approaches to achieve quantum supremacy for practical problems. Typically, VQA is a hybrid quantum-classical algorithm that uses classical optimizer to find the parameters $\theta$ that minimize the cost function $C(\theta)$.

\begin{equation}\label{theta_define}
\theta^{*} = \underset{\theta}{argmin} \  C(\theta).
\end{equation}

The key idea of VQAs is to encode the problems into a cost function such that the minimum of the cost function corresponds to the solution of the problem. In general, the cost function can be defined as:

\begin{equation}\label{cost_func}
\theta \space=\space \sum_{k}f_{k}(\{ \rho_{k} \},\{ O_{k} \},\{ U(\theta) \})
\end{equation}

\noindent 
where $f$ is a function, $U(\theta)$ is the parameterized circuits (ansatz), $\rho_{k}$ represents inputs from the training set, $O_{k}$ are the observables for measurements.

For example, in quantum chemistry, the variational method is a classical method to find low energy states of the quantum system. A trial wave function (ansatz) is defined with parameters. The expectation of the energy changes as the parameters vary. Then we can use classical optimizer to search for the minimum energy. The minimized ansatz is the approximation of the lowest energy states, and the corresponding energy gives an upper bound on the actual ground energy. In QML problems, however, the resulting minimized ansatz does not have physical meaning.
It leads to a larger space to design the QML ansatz. 
% Figure \ref{fig_base} shows different types of ansatz that can be adopted for QNN algorithms.

\subsection{Quantum Neural Network}

QNN is a subset of VQAs that is composed of encoder, ansatz, decoder, etc. Abbas {\em et al.}~\cite{abbas2021power} claims that the well-constructed QNNs can have substantially higher effective dimensions than classical neural networks, meaning that they can model a broader class of functions. 
In another word, QNN can achieve better 
performance than the comparable classical feedforward neural networks. Importantly, the benefit comes without the cost of reduced trainability. Figure~\ref{fig_qnn} shows the basic structure of the QNN programs. The quantum circuits used in QNN contains an encoder and an ansatz, i.e., variational quantum circuits with trainable parameters. The input features are encoded inside the encoder gates as phase or amplitudes of rotations. These gates act as preliminary part for the ansatz. After the QNN circuits are combined, the measurements are performed, and we can obtain an output distribution. The value of the cost function is updated according to the output distribution. Gradients of the parameters are calculated with the parameter shift rule~\cite{crooks2019gradients} and the trainable parameters are updated accordingly. 

\subsection{Topology of Quantum Hardware}

\begin{figure}[t]
\centering
\includegraphics[width=\linewidth]{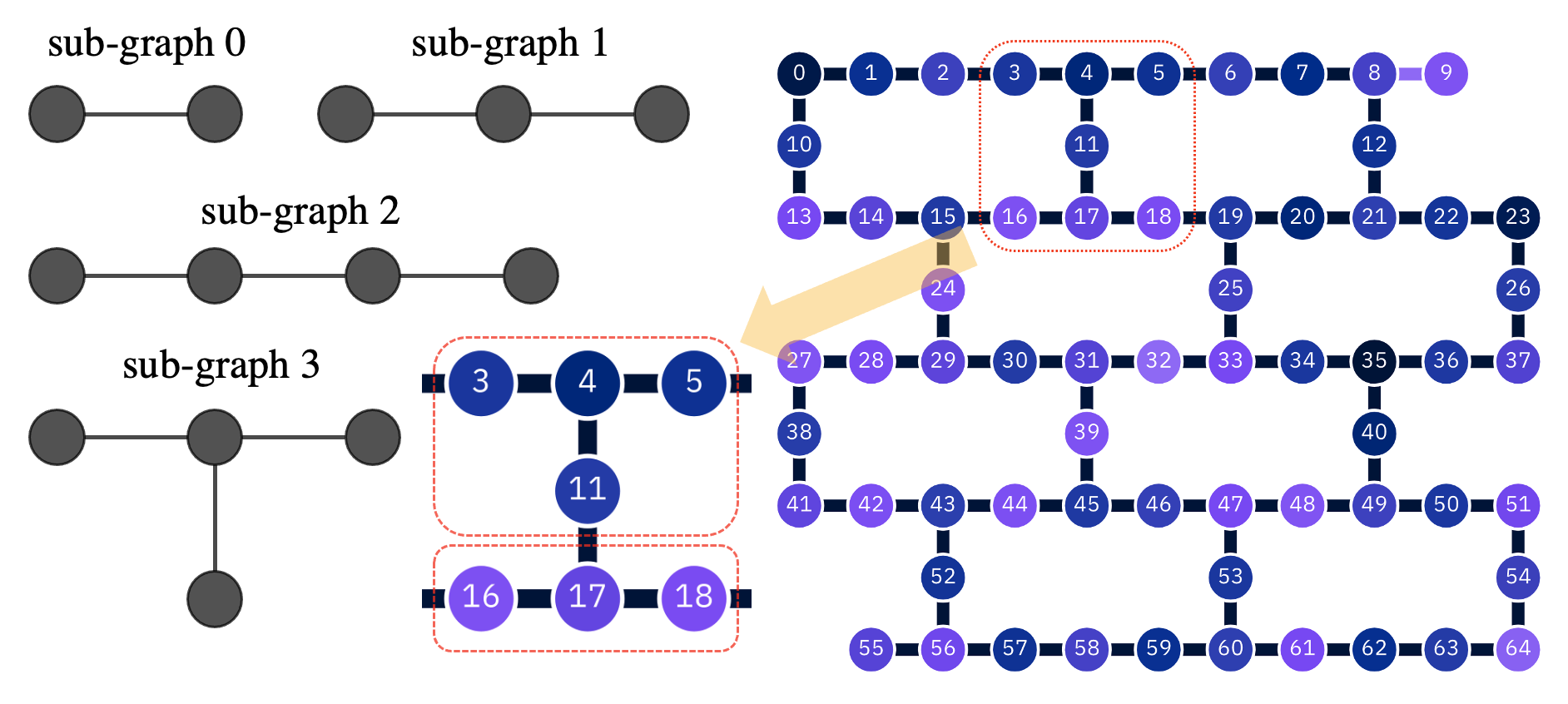}
\caption{Hardware topology of a real-world quantum computer ibm\_ithaca~\cite{ibm_device}. 
% The quantum machine has 65 qubits and 72 connections between them in total. 
The topology information is considered when sub-circuits are generated based on 4 sub-graphs shown in the figure.}
\label{fig_ibm_topo}
\end{figure}

In currently available quantum devices, the quantum bits are usually not fully connected. The quantum bits are often arranged as 1-dimensional array or 2-dimensional grid-like or heavy-hexagon arrays. The connection between quantum bits is actually very sparse compared with the full connection. 
% For fully connected quantum computers with $N$ qubits, the number of connections is $O(N^2)$. However, in the currently available quantum devices, the number of qubit connections is $O(N)$. 
For example, ibm\_ithaca is a 65-qubit quantum machine, but the total number of connections is 72 as shown in Figure \ref{fig_ibm_topo}. 
% The connections between qubits will get sparser for larger quantum computers in near future. 
% In this case, a large 
% portion of two-qubit gates in a quantum program
% cannot be directly executed. 
As a result, when the quantum programs are executed 
on the quantum machines, a considerable number of SWAP gates are inserted
to the original quantum circuit to
compensate for the sparse connection. We tested several 4-qubit quantum programs on the real-world hardware topology ibmq\_quito \cite{ibm_device}.
We found that the mapped program has around 30\% more gates, and the circuit depth is increased by 40\%. Such overheads vary from programs to programs. 
In general, we expect that 
the number of extra SWAP gates will become larger in near future, due to the increasing sparsity of physical qubit connections.

% \clearpage

\section{Proposed Bottom-up Approach}
\label{over}

\begin{figure*}[h!tb]
\centering
\includegraphics[width=0.9\linewidth]{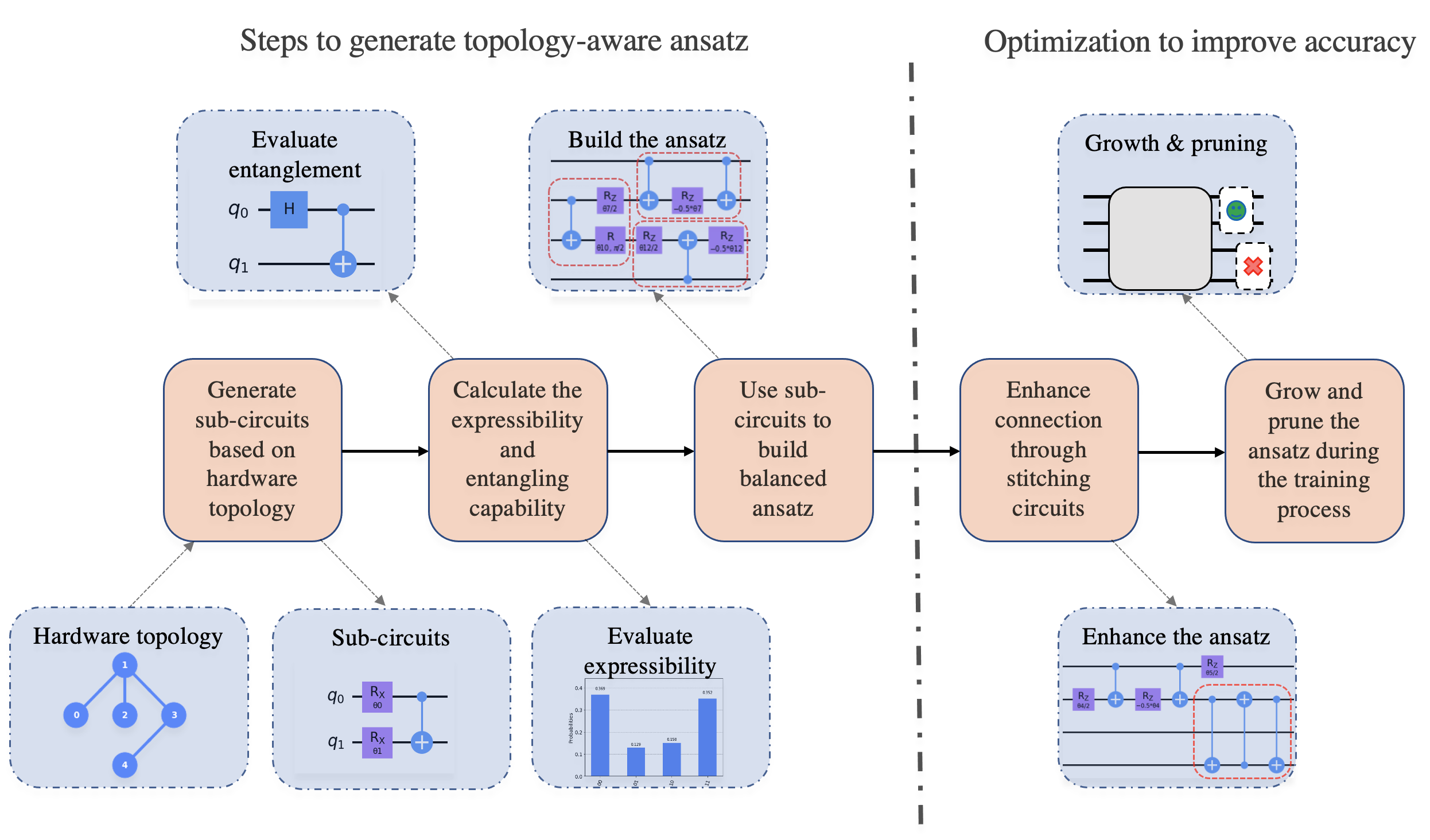}
\caption{Overview of the Proposed Approach: (1) Sub-circuits are generated based on hardware topology; (2) The circuits are evaluated with two criteria; (3) Sub-circuits are used as building blocks to generate an initial ansatz; (4) After the sub-circuits are concatenated together, additional two-qubit gates are added to boost the entanglement; (5) Ansatz is appended with small sub-circuits
and trained to further improve accuracy.}
\label{fig_overview}
\end{figure*}

\subsection{Motivation}

%%Motivation, why methodology can help QNN more, because the circuits are not fixed before the synthesis, the QNN provides flexibility to change the structure of the circuits. Why need sub-circuits, better properties, consideration of topology, highest level motivation? Challenge of the large search space for the final ansatz, building block of the final ansatz, building block requirement: compatible with topology,%%

With the current approaches, the mapping process is performed 
after the quantum circuits are synthesized.
The quantum hardware topology is not considered during synthesis.
As discussed earlier, it introduces additional overhead due to 
the inserted SWAPs, and affect the selection of proper 
ansatz for NAS. 
Intuitively, considering hardware topology during synthesis
has the potential to generate better ansatz with less circuit depth.
It is not possible with the current top-down NAS approach that 
generates the ansatz candidates from a topology-oblivious
super-circuit. 
A straightforward solution is to generate the ansatz
from scratch, and ensure that each generated ansatz is compatible
to the hardware topology.
However, this approach will lead to huge design space and 
requires very high computational resource to search for the ansatz. 
Conceptually, there is a {\em gap} between generated 
ansatz and hardware topology. 

\subsection{Sub-circuit as Building Blocks}

To close this gap and make the optimization problem tractable, we propose \name, a {\em bottom-up approach using sub-circuits with compatible 
topology as the building
blocks} to generate the ansatz for QNN.
For a given quantum computer, the ansatz can be generated by combining the
set of compatible sub-circuits---the sub-graphs corresponding
to the sub-circuits can be embedded in the graph corresponding
to the hardware topology. 
By design, the combinations of the sub-circuits require 
no extra SWAP gates in the mapping process.
In our approach, the search of a huge design space is reduced to 
finding better sub-circuits that can form the high quality ansatz.

We use two criteria to evaluate the sub-circuits based on their capabilities to explore the Hilbert Space.
The circuits are ranked and grouped according to their size and performance.
The details are discussed in Section~\ref{eval}. 
Then the sub-circuits can be combined together to form
an initial QNN ansatz.
Since it is composed of multiple independent blocks, 
further optimizations are likely to be achievable. 
We propose several methods to boost the performance.
First, after the sub-circuits are combined, 
we can add more two-qubit gates to compensate for the 
connection loss between the qubits.
Second, during the training process, 
we ``grow'' the ansatz by appending small sub-circuits. 
This step terminates when no further reduction of the cost function
is observed.
Third, we adopt the idea of dynamic pruning~\cite{sim2021adaptive} 
to further reduce the size of the ansatz.
Specifically, the gates with parameters that are close to zero
are deleted. 
This step terminates if a sharp decrease of model accuracy is observed.
The details of the search and optimization process 
are discussed in Section~\ref{opt}. 

In our approach, the reduction in search space may 
sacrifice some opportunity to find the optimal solution. 
However, given the huge search space for a bottom-up approach,
the trade-off is worthwhile to make the optimization problem tractable.
In our evaluation, QNNs are constructed and tested on various learning problems. The results show that our approach
can generate QNN ansatz with similar---sometimes even better---accuracy
compared to the state-of-the-art solutions with significantly
reduced circuit size. 
Specifically, under similar performance, we achieved 
around 50\% reduction on circuit depth and up to 75\% reduction on the number of CNOT gates.

%We notice that normally the mapping process happens after the quantum circuits are synthesized. Namely, the information of hardware topology is omitted in the process of synthesis. We believe that the extra information from the hardware topology could help us generate better ansatz of less circuit depth. However, the design space is large. If we don't divide the circuits into smaller sub-circuits, we will need huge computational resource to search for the best ansatz. To make the optimization problem tractable, we propose a bottom-up approach, where sub-circuits are optimized and then combined to generate the ansatz. In such manner, the design space is reduced to find the sub-circuits that will form the better ansatz. After the sub-circuits are merged to form the ansatz, further optimization will be applied to compensate for the connection sparsity.

\subsection{Why Does It Work?}

Classical machine learning and variational algorithms in quantum mechanics share similar mathematical structure~\cite{stoudenmire2016supervised}. Many natural quantum systems satisfy the area law of entanglement, which implies that the entanglement entropy scales as the surface area of the subsystem rather than its volume~\cite{eisert2010colloquium}. For example, the {\em ground states} of many typical Hamiltonians satisfy the area law~\cite{sarma2019machine}, indicating that the relevant physics only takes place in a restricted part of the full Hilbert space. Our solution is
inspired by the phenomenon of the area law: it is possible that we do not
need to search in the full Hilbert space as QNN ansatz. 
Our solution explores such ``locality'' with the bottom-up approach.  
Our experimental results indeed 
suggest that local optimizations of sub-circuits can help improve the global performance.

%Traditionally, ansatz are designed specially for different applications. For instance, the hardware-efficient ansatz~\cite{kandala2017hardware} and the famous UCCSD~\cite{barkoutsos2018quantum,sokolov2020quantum} are specifically designed for VQE algorithms. In the case of QNN algorithms, it remains an open question on how to find the optimal ansatz. Our goal is to eliminate the need of SWAP gates for QNN ansatz, while maintaining the model accuracy. The development of quantum compilers~\cite{murali2019noise,chong2017programming} has made the synthesis and mapping process very efficient. To further reduce the size of the ansatz, we 
%propose to consider hardware topology during synthesis. QNN has flexibility on how circuits are formulated. This property grants us space to optimize the ansatz design during synthesis.

%The optimization problem comes with a large search space. We propose to generate the ansatz in a bottom-up approach, so that the search space for these sub-circuits are controllable. The reduction in search space will sacrifice some probability to find the optimal solution. But the trade-off is worthwhile to make the optimization problem tractable. \violet{We will first generate sub-circuits from the information of hardware topology. These sub-circuits are directly compatible with the hardware topology and will act as building blocks for the final ansatz. In this way, the combinations of these sub-circuits require no extra SWAP gates in the mapping process. }

% \clearpage

\section{Sub-circuit Selection}
\label{bua}

%\textcolor{blue}{Addition of design space explanation}

\subsection{Generating Sub-circuits}

\begin{algorithm}[t]
\caption{Procedure to generate a sub-circuit}  
\label{sub_gen}  
\begin{algorithmic} [1]
\REQUIRE Topology information, limit on circuit depth
% \ENSURE large-groups
\STATE Check available two-qubit connections from topology information and generate a set of compatible gates
\STATE The set contains 1-qubit gates on all available qubits and 2-qubit gates on all available connections
% \STATE Determine the gate set
\WHILE{Circuit depth $\leq$ depth limit}
\STATE Select a random gate from the set
\IF{The gate is not the same gate with the last gate}
\STATE Assign the gate to the qubit with least gates
\ENDIF
\STATE Update \#gates on each qubit
\ENDWHILE
\STATE Set gate parameters to zeros
\IF{The resulting Unitary equals identity matrix}
\STATE The sub-circuit is labeled as appendable
\ENDIF
\RETURN One sub-circuit
\end{algorithmic}  
\end{algorithm}

%\textcolor{cyan}{
%Q: Why expressibilty and entangling capability as main criteria?
%}
%\textcolor{red}{
%Q: Why only four sub-graphs are selected to generate the ansatz?
%}

%\red{NEED TO REWRITE THIS PARAGRAPH AND EXPLAIN WHY THESE SUB-CIRCUITS CAN COVER ALL? TOPOLOGY PARTITION. SUBGRAPH, ETC. ALSO NEED TO UPDATE FIGURE 6. QNN ansatz are generated from small blocks of sub-circuits. }
%The properties of the sub-circuits are measured according to the criteria described above.

The SWAP gates are needed whenever the topology of a quantum circuit is different from the quantum computer topology. 
Typically, the connections between physical qubits are sparse for superconducting quantum computers, as shown in the 
example of Figure~\ref{fig_ibm_topo}. 
For a given quantum computer, we consider the hardware topology as a connected graph, where qubits are vertices and available connections are edges. 
For the ibm\_ithaca, 
we select 4 sub-graphs as bases to make our ansatz. 
Since the smallest sub-graph 0 contains only 2 qubits, only single isolated qubit can be uncovered in the worst case. For example, we want to design a 7-qubit ansatz. We can first select the 7 qubits from the quantum hardware, then divide the 7 bits into sub-graphs of size 3 and 4, as shown in Figure \ref{fig_ibm_topo}. If any uncovered single qubit exists, we can apply stitching circuits on it to connect isolated sub-graphs.
% , more details are described in Section \ref{stitch}.

Based on the allowed sub-circuit topologies, 
we can generate small blocks of variational circuits that are 
compatible with the native quantum hardware topology.
%We first define small blocks of variational circuits. These circuits are generated in a way that they are natively compatible with the quantum hardware topology. 
Such circuits are small in the aspect of width, depth and the number of parameters. The gate set for sub-circuits to choose from is $\{CNOT, R_{x}(\theta), R_{y}(\theta), R_{z}(\theta)\}$. The gate set can be extended to include \{$\sqrt{X},H$\} gates as well. The controlled rotation gates can be decomposed into these gates, so it is excluded from the gate set. We also notice that $R_{x}$ and $R_{y}$ gates can be further decomposed into $R_{z}$ gates and $\sqrt{X}$ gates. We decide to include the single-qubit rotation gates, since their combinations can boost the expressibility better than random combination of $R_{z}$ gates and $\sqrt{X}$ gates. 

With the topology information from quantum device, we can easily form a set of compatible gates. For example, if the quantum computer 
has the connection between qubit 0 and qubit 1, then compatible gates are 1-qubit gates and 2-qubit gates on qubit 0 and 1. To be more specific, 1-qubit gates are $\{R_{x}(\theta), R_{y}(\theta), R_{z}(\theta)\}$ on both qubits; 2-qubit gates are $\{CNOT(0,1), CNOT(1,0)\}$. Therefore, the two-qubit gates are assigned only to available physical qubit connections. We notice that two identical rotation gates are the same as one rotation gate with parameter that is the sum of two separate parameters; and two consecutive CNOT gates form an identical matrix. Consequently, when compatible gates are assigned onto qubits, consecutive identical gates will be omitted. 
% Figure \ref{fig_ex} demonstrates a simplest case where the sub-circuits contain only two qubits. The need for SWAP gates is by design unnecessary for these sub-circuits. 

The algorithm to randomly generate
sub-circuit is shown in Algorithm~\ref{sub_gen}.
After generated, the sub-circuits are ranked with their performance (to be defined in the next section). The top sub-circuits are saved in the library for constructing QNN ansatz. During the process of sub-circuits generation, we divide the sub-circuits into two groups:
the circuits that result in identity matrix with the parameters set to zeros; and
the rest ones. 
The reason for grouping is to identify the sub-circuits that are suitable for proposed optimization method,
more details
will be discussed in Section~\ref{growth}.

\subsection{Sub-circuits Quality Criteria}
\label{eval}

After the sub-circuits are generated, we measure the properties of them using the 
{\em expressibility} and the {\em entangling capabilities}. These properties are chosen as main criteria since~\cite{hubregtsen2021evaluation} demonstrates that they are highly correlated with the model accuracy. We expect to see better model accuracy with ansatz of better properties.

%The expressibility generally measures the ability of a circuit to uniformly address the Hilbert space. And the entangling capabilities quantifies the ability of a circuit to capture non-trivial correlations in the quantum data. These sub-circuits will be stored in a library with the measured properties. ~\cite{hubregtsen2021evaluation} points out that expressibility/entangling capability and model accuracy are correlated. In subsection \ref{exp} and subsection \ref{ent}, the two criteria will be discussed in details.

%\subsection{Expressibility}
%\label{exp}

% \begin{figure}[h!tb]
% \centering
% \includegraphics[width=0.7\linewidth]{Figure/bloch_0_3.png}
% \caption{Expresssibility measures a circuit's capability to generate quantum states that can fully explore the Hilbert space. }
% \label{fig_bloch}
% \end{figure}

\begin{figure}[t]
\centering
\includegraphics[width=\linewidth]{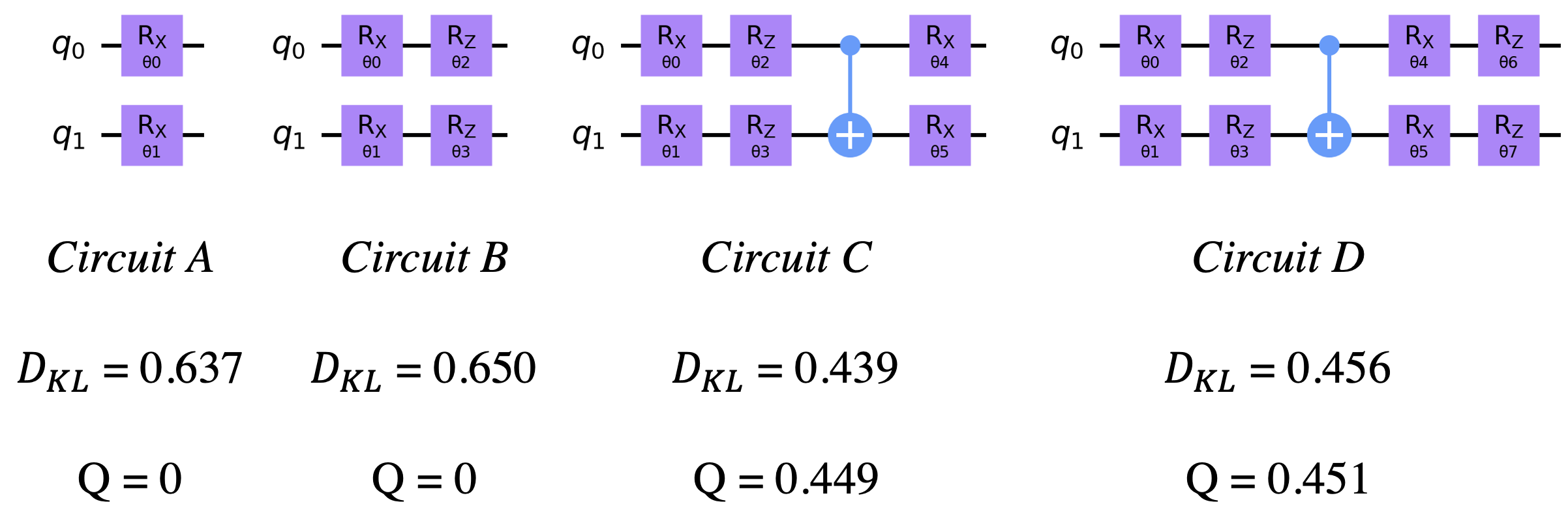}
\caption{Sub-circuit examples: from left to right, the circuits are larger with better expressibility and entangling capability.}
\label{fig_ex}
\end{figure}

{\bf Expressibility} 
It is defined as a circuit's ability to generate states that well represent the Hilbert Space~\cite{sim2019expressibility}. In the case of single-qubit system, it can be interpreted as how uniformly the Bloch Vectors are distributed on the Bloch sphere.
% as shown in Figure \ref{fig_bloch}. 
To quantify the expressibility, we use the Kullback-Leibler (KL) divergence~\cite{kullback1951information} to measure the difference between two distributions:
$Expr\space=\space D_{KL}(\hat{P}_{ansatz}(F;\Theta)||P_{Haar}(F))$,
where $\hat{P}_{ansatz}(F;\Theta)$ describes the distribution of estimated fidelities $F\space=\space |\langle \psi_\theta|\psi_\phi\rangle|^2$ with randomly sampled parameter pairs $(\theta,\phi)$. $P_{Haar}(F)$ stands for the uniform distribution of states. $P_{Haar}(F)$ can be derived in theory. Therefore, in experiments we only need to measure the distribution of $\langle \psi_\theta|\psi_\phi\rangle$.

Figure~\ref{fig_ex} shows examples 
of sub-circuits
computational expressibility. For circuit A, only $R_{x}$ gates are applied, the degree of freedom is very limited. Thus, the fidelity distribution will be very different from the uniform distribution. To quantify the difference, KL divergence is applied. For circuit C, where different rotation gates and CNOT gates are applied, the quantum states will explore the Bloch spheres more thoroughly. The distribution of the fidelity should be closer to the uniform distribution. Hence, we expect to see better expressibility for ansatz with more parameterized gates. On the other side, the sheer 
inclusion of parameterized quantum gates can lead to problems like barren plateaus~\cite{mcclean2018barren} and trainability~\cite{thanasilp2021subtleties}. 
Thus, we need to limit the size of the ansatz to avoid such problems.

%\subsection{Entangling capability}
%\label{ent}

{\bf Entangling capability}
For QNN algorithms, the solution space for data classification tasks needs to be efficiently represented. In this context, the entangling capability provides potential advantages in capturing the non-trivial correlations in the datasets ~\cite{sim2019expressibility}. During the process of generating the sub-circuit library, the entangling capability is quantified by the Meyer-Wallach (MW) entanglement measure~\cite{meyer2002global}. It is a global measure of multi-qubit entanglement for quantum states. While there are other methods for quantifying the entanglement, MW is generally more scalable and easy to compute:
$Q(|\psi\rangle)=2(1-1/n\sum_{k=0}^{n-1}Tr[\rho_{k}^2])$,
where $\rho_{k}$ is the one-qubit reduced density matrix of the k-th qubit after tracing out the rest~\cite{brennen2003observable}. 
The values of Q range from zero (no entanglement) to one (strong entanglement). 
This equation illustrates the physical meaning of such measure: it is an average over the entanglements of each qubit with the rest of the quantum system. If we measure the MW for bell state, $|\psi\rangle = \frac{1}{\sqrt{2}}(|00\rangle+|11\rangle)$, we will obtain $Q=1$. 

\begin{algorithm}[t]
\caption{Combination of sub-circuits}  
\label{sub_comb} 
\begin{algorithmic} [1]
\REQUIRE Hardware topology, evaluated sub-circuits
% \ENSURE large-groups
\STATE Select sub-circuits of best expressibility or best entanglement to build the ansatz 
%\STATE Can also select sub-circuits of best entanglement
\WHILE{Circuit depth $\leq$ depth threshold}
\STATE Assign the sub-circuits to the physical qubit with least gates or assign the sub-circuits in the order of decreasing size
\STATE Update \#gates on each qubit
% \STATE Can also 
\ENDWHILE
\RETURN An initial QNN ansatz
\end{algorithmic} 
\end{algorithm}
% \vspace{-3mm}

To evaluate the entanglement capability, multiple sets of random parameters are applied to the sub-circuit, and the corresponding $Q$ values are calculated. We define the MW entanglement for the sub-circuit as the average value of $Q$.
In Figure~\ref{fig_ex}, for circuit A and B, different qubits do not interact (not entangled). The MW measure will return 0 for such situations. For circuit C and D, the two qubits are entangled through CNOT gates. Because they are parameterized circuits, the value of $Q$ changes as parameters change. The displayed values are calculated by averaging a large number of quantum circuits with randomly sampled parameters. The MW entanglement measure has one drawback: it cannot distinguish two quantum states with very high entanglement because 
the MW entanglement measure will saturate. 
Fortunately, this drawback does not affect our evaluation of the sub-circuits. 
Because we calculate the {\em average} of $Q$ values for different sets of parameters, although it is possible that $Q=1$ for {\em one} set of parameters,
it is highly unlikely that $Q=1$ for {\em all} possible sets of parameters.
% The saturation of MW entanglement can be reached for {\em one} set of parameters. 
% It is unlikely that the sub-circuits will result in full-entanglement with {\em all} possible parameters.
Later in Figure~\ref{fig_cri}, we can see that the MW measure of some ansatz are 0, but it is almost impossible that we experience entanglement saturation, i.e., obtaining 1 for MW measure. We notice that the Q values will saturate with more CNOT gates, which actually fits our principles. Since we want to limit the size of the circuits to narrow down the design space.

% \clearpage

\section{Searching High Accuracy Ansatz}
\label{opt}

\subsection{Combining Sub-circuits}
\label{comb_sub}

\begin{figure}[t]
\centering
\includegraphics[width=\linewidth]{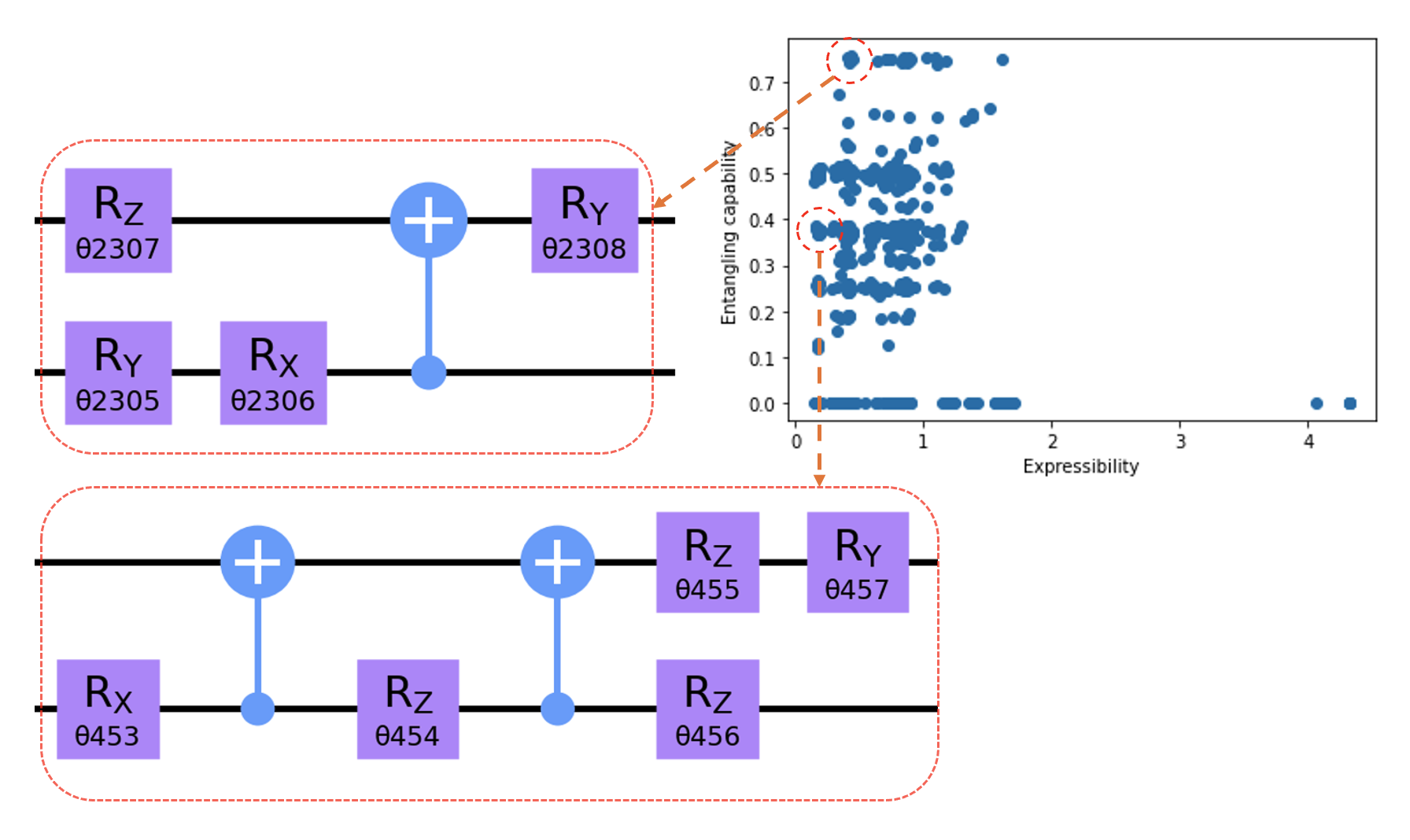}
\caption{Two sub-circuits are selected according to their great expressibility and entangling capability respectively. The combination generates a 4-qubit ansatz with 12 gates.}
\label{fig_small_ex}
\end{figure}

After obtaining a library of sub-circuits, we combine them to produce an initial ansatz. We give options to combine the circuit, for example, we can assign the sub-circuits onto the compatible qubit with least gates. The consequent ansatz would be balanced in terms of gates per qubit. We also allow to assign the sub-circuits in the order of their sizes. In this case, the ansatz is similar to traditional neural network with layers of decreasing dimensions. We observed 
negligible accuracy change among different combination methods. 
Thus, we choose the simple method to form a balanced initial ansatz. 
The assignment of sub-circuits ends when a threshold of depth is reached. The details are described in Algorithm \ref{sub_comb}. We demonstrate an example in Figure \ref{fig_small_ex}.

\subsection{Stitching Sub-circuits} \label{stitch}

\begin{figure}[t]
\centering
\includegraphics[width=\linewidth]{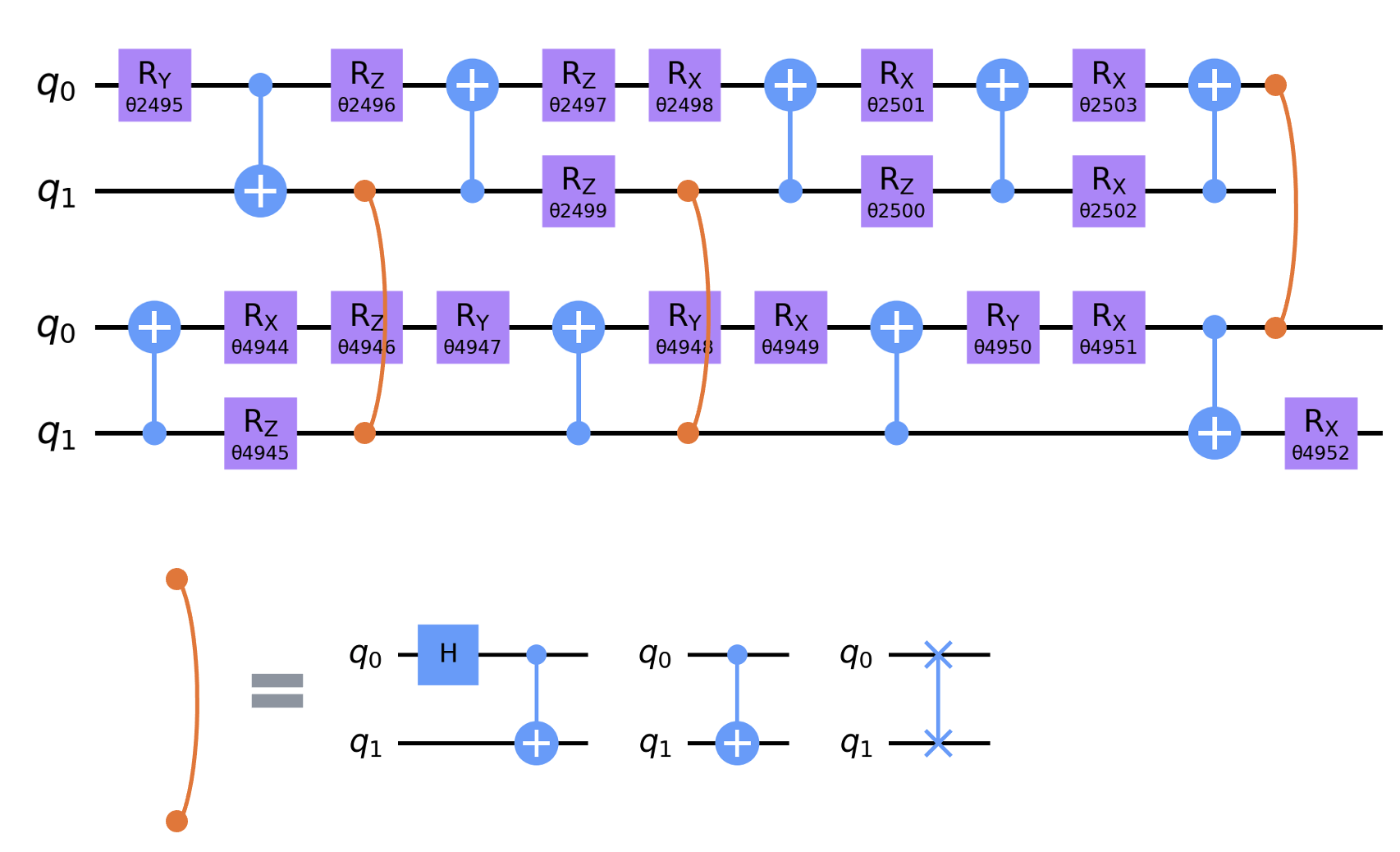}
\caption{We propose to insert two-qubit gates to mitigate the problem that sub-circuits do not interact with each other. Illustrated here are the possible options.}
\label{fig_stitch}
\end{figure}

Since the sub-circuits in the library are limited in terms of depth and width. The expected global entanglement of the initial ansatz is not strong. 
As a result, the performance of the topology-aware ansatz may not reach the expectation. To fix this problem, we propose to use ``stitching circuits'' to boost the performance of the entire circuit. The stitching circuits are two-qubit gates that can enhance the entanglement. \{CNOT,CRX,SWAP\} and other two-qubit circuits can serve as stitching circuits. We apply the stitching circuits to places where the qubit 
is idle or at the end of the sub-circuits, as shown in Figure \ref{fig_stitch}.

\begin{figure}[t]
\centering
\includegraphics[width=\linewidth]{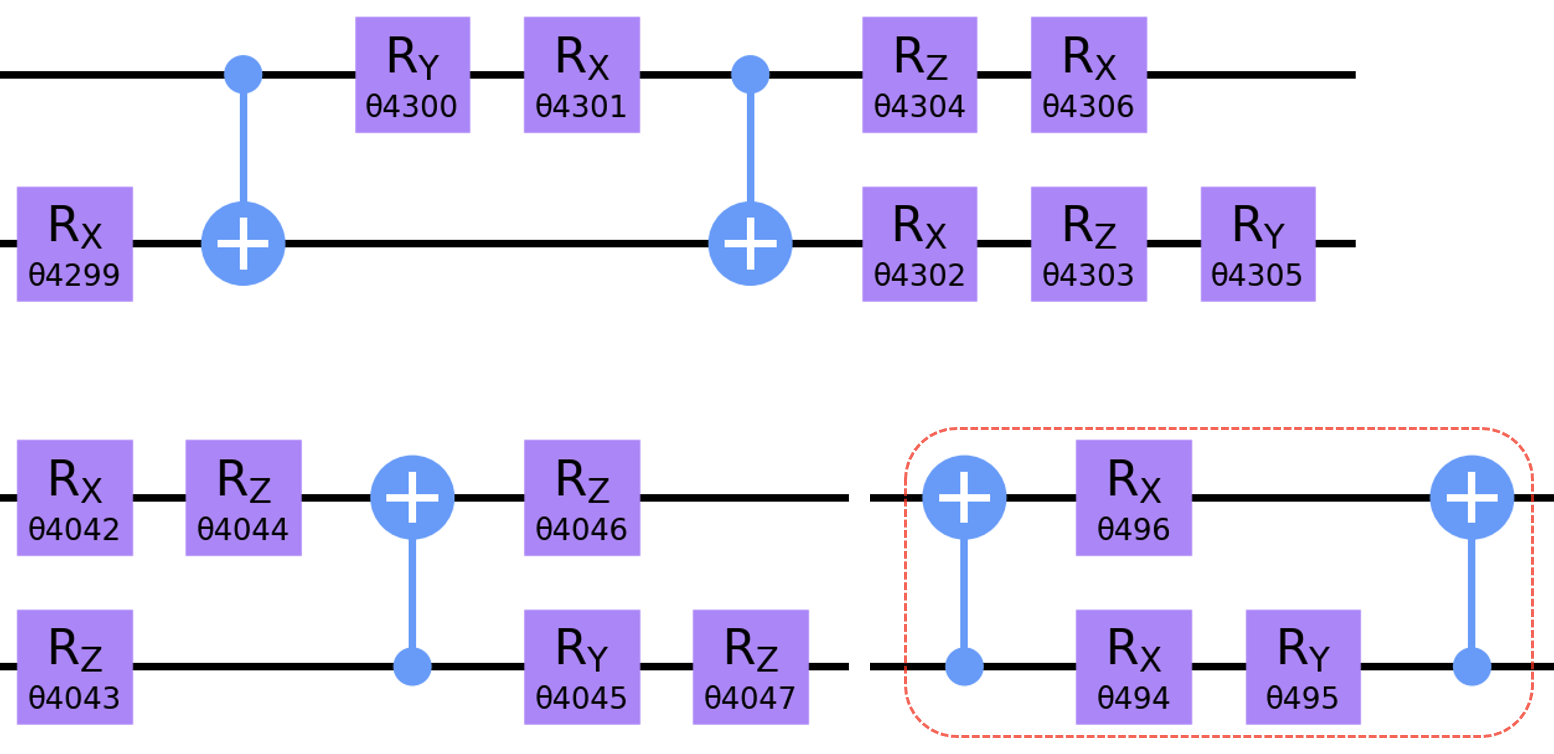}
\caption{During the training process, we propose to insert small sub-circuits at the end of the ansatz. 
The unitary matrix of this small ansatz is identity matrix if the parameters are initialized to zeros. Hence, the gradients of all the other parameters will not be abruptly changed.
}
\label{fig_growth}
\end{figure}

\begin{figure}[h!tb]
\centering
\includegraphics[width=\linewidth]{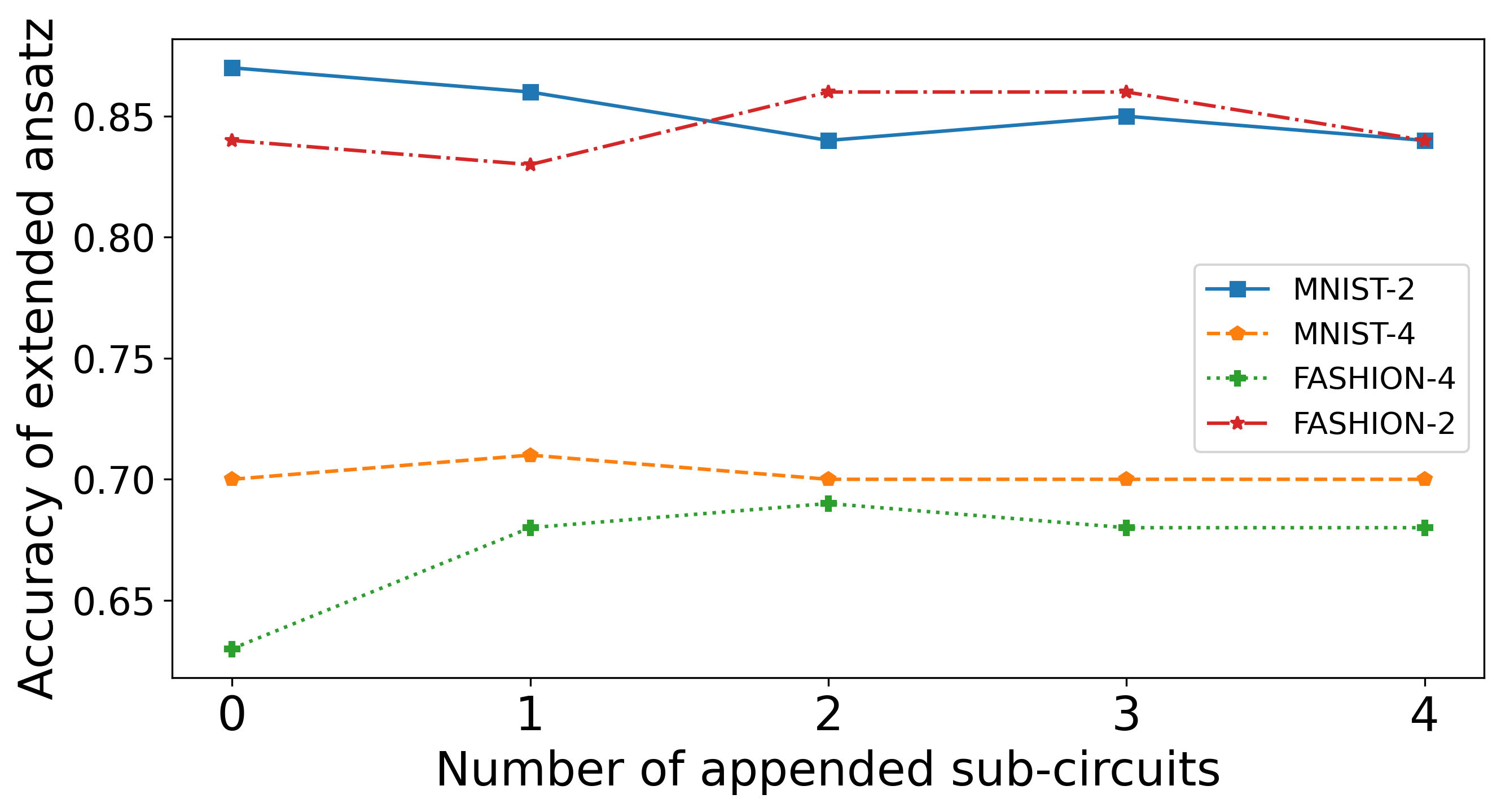}
\caption{The model accuracy varies with different numbers of appended sub-circuits. }
\label{grow_accu}
\end{figure}

\subsection{Growing Circuit During Training}\label{growth}

As shown in Algorithm~\ref{sub_gen}, sub-circuits are classified into two groups. 
The sub-circuits in one group have an identity unitary matrix when parameters are set to zeros, and the rest form the other group. The purpose of grouping is to dynamically ``grow'' the ansatz if it has inferior performance. In the process of training, we append the sub-circuits from the first group to ansatz. Since they are identity matrices in the beginning, the training loss of the appended ansatz will be smooth without abrupt changes. In this way, we can tell if the incremental addition is effective by observing the training loss in the first few iterations. 
This strategy allows us to refine the ansatz and boost its performance incrementally. 
The growing process ends if the model accuracy is not improved. 
Figure~\ref{fig_growth} provides an example, in which
the sub-circuit inside the rectangular is appended to the ansatz. 
Figure \ref{grow_accu} shows how accuracy changes 
with the number of appended sub-circuits. 
For different data sets (details in Section~\ref{eva}), we can see that 
there is always a sweet spot where the ansatz capacity is improved, 
while the optimizer can still handle the additional parameters.
We observe a slight drop of accuracy as the number of appended sub-circuits further increases. 
We hypothesize that it is because
the optimizer cannot handle too many additional parameters with limited iterations.

\subsection{Dynamic Gate Pruning}\label{dy_prune}

\begin{figure}[t]
\centering
\includegraphics[width=\linewidth]{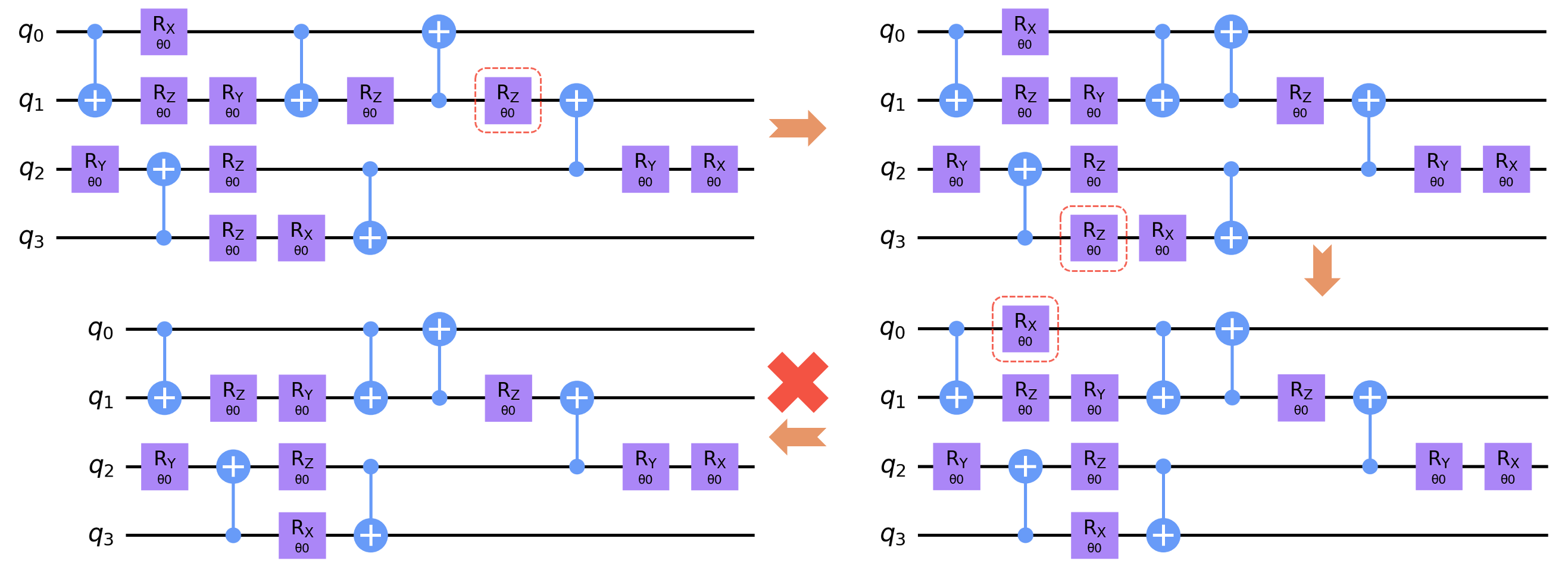}
\caption{Dynamic gate pruning will remove the gates with parameters that are close to zeros. The model accuracy for these ansatz are [0.80, 0.81, 0.80, 0.75]. The removal of the first two gates does affect the model accuracy, while the removal of the last gate results in obvious accuracy drop.}
\label{fig_prune}
\end{figure}

During the training process, some parameterized gates can be pruned according to the absolute values of the parameters. If the parameter of the gate is close to zero, we will prune the parameterized gate and monitor the change in accuracy. The pruning process ends if a sharp drop of accuracy is observed. A trade-off exists between the pruning and the overall accuracy. Figure~\ref{fig_prune} illustrates an example of the ansatz before and after pruning. We can see that two out of 11 rotation gates are removed from the original ansatz. The gate pruning terminates at this point since further reduction of parameterized gates would result in a 5\% decrease of accuracy (from 80\% to 75\%). In comparison, the removal of the first two parameterized gates leads to no reduction in model accuracy. Even though the ansatz is small, it achieves a classification accuracy of 81\% for the two-class MNIST problem. It validates the great potentials of QNN algorithms: quantum circuits with strong capability to represent data.

% \begin{figure}[h!tb]
% \centering
% \includegraphics[width=0.9\linewidth]{Figure/datasets_0_0.png}
% \caption{The datasets and their figures from ~\cite{hubregtsen2021evaluation}. Each dataset contains 1500 data points.}
% \label{fig_data}
% \end{figure}

% \clearpage

\section{Evaluation}
\label{eva}

\subsection{Datasets and Backend Configuration}

There are many datasets for the evaluation of classical machine learning algorithms. However, such datasets are usually large and too complicated for QNN algorithms. To evaluate the QNN circuits that are generated with our bottom-up methods, we adopt the datasets used in ~\cite{hubregtsen2021evaluation} and ~\cite{wang2021quantumnas}. The datasets from ~\cite{hubregtsen2021evaluation} is composed of nine tasks of increasing difficulty but of suitable size. Each dataset contains a total of 1500 data points for training, testing and validation. 
% The visualization of datasets is shown in Figure \ref{fig_data}. 
The datasets from ~\cite{wang2021quantumnas}, on the other hand, are adopted from real-world applications such as MNIST~\cite{deng2012mnist}, FASHION-MNIST~\cite{xiao2017fashion} and VOWEL~\cite{Dua:2019}. The input images from MNIST and FASHION are cropped to $24\times24$ and down-sampled to $4\times4$ with average pooling. Inputs from VOWEL are reduced to contain 10 features. Measurements of ansatz are conducted on Pauli-z basis to obtain the expectation values for each qubit. Then softmax is applied to produce probability values for different classes.
As for backends, we use torch quantum~\cite{wang2021quantumnas} to perform simulations on classical computers. Even though our ansatz are generated based on the topology of IBM's quantum devices~\cite{ibm_device}, they are fine-grained enough to fit other kinds of backends such as trapped ion quantum computers. Experiments on real-world quantum computers are conducted on IBM's quantum computers such as ibmq\_quito and ibmq\_lima. Further details will be discussed in subsection \ref{real}.

\subsection{Sub-circuits Generation}

\begin{figure}[t]
% \vspace{-4mm}
\centering
\includegraphics[width=\linewidth]{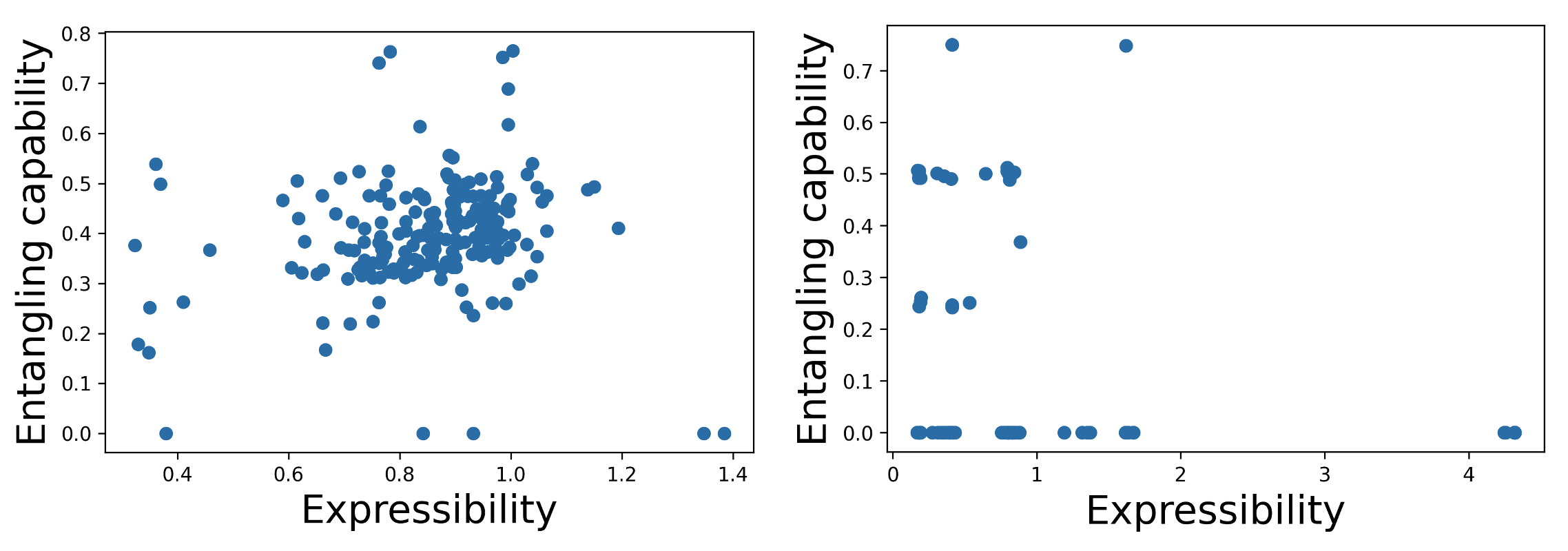}

\caption{The scatter plots show the relationship between expressibility and entangling capability. The right plot is generated for sub-circuits with less than five gates, while the sub-circuits in the left plot contain around 10 gates each. If the sub-circuits are too small, the search space for ``good'' ansatz will also be very limited.}
\label{fig_cri}
\end{figure}

The first step of our bottom-up approach is to create a large number of sub-circuits and evaluate them. Some example circuits and the values of corresponding criteria are shown in Figure \ref{fig_ex}. When we generate the sub-circuits, the size of the sub-circuits is limited. If the sub-circuits are too small, they cannot thoroughly explore the Hilbert Space with varying parameters. Figure \ref{fig_cri} demonstrates the situation where small sub-circuits result in minimum search space. On the other hand, if the sub-circuits are too large, it obeys our principles to narrow down the design space.

\begin{table*}[h!tb]
% \vspace{-7mm}
\centering
\caption{Ansatz size comparison}
\resizebox{\textwidth}{!}{
\begin{tabular}{cccccccc}
\hline
                       & Depth  & \#Gates  & \#Params & Avg accuracy & Compiled depth & Compiled \#gates & Compiled \#CNOT gates \\ \hline
Base\_1   & 5      & 11       & 8        & 0.54             & 12                      & 30                        & 3            \\ \hline
Base\_2   & 6      & 19       & 19       & 0.56             & 26                      & 66                        & 6            \\ \hline
Base\_3   & 5      & 11       & 4        & 0.61             & 20                      & 40                        & 3            \\ \hline
Base\_4   & 6      & 15       & 12       & 0.72             & 18                      & 39                        & 3            \\ \hline
Base\_5   & 9      & 16       & 16       & 0.74             & 49                      & 83                        & 16           \\ \hline
Base\_6   & 9      & 16       & 16       & 0.74             & 95                      & 140                       & 34           \\ \hline
Base\_7   & 9      & 16       & 8        & 0.51             & 31                      & 55                        & 16           \\ \hline
Base\_8   & 6      & 12       & 12       & 0.64             & 54                      & 77                        & 17           \\ \hline
Base\_avg & 6.9    & 14.3     & 11.9     & 0.63             & 38                      & 66.3                      & 12.3         \\ \hline
Initial ansatz         & 6      & 12       & 9        & 0.73             & 21                      & 34                        & 3            \\ \hline
Stitched ansatz          & 6      & 13       & 9        & 0.71             & 18                      & 32                        & 4            \\ \hline
Grown ansatz         & 6 to 7 & 13 to 17 & 9 to 12  & 0.68             & 18 to 22                & 32 to 45                  & 4            \\ \hline
Pruned ansatz          & 7      & 13 to 15 & 9 to 10  & 0.7              & 18 to 20                & 32 to 40                  & 4            \\ \hline
\end{tabular}
% \vspace{-8mm}
}
\label{tab_1}
\end{table*}

\begin{figure*}[t]
\centering
\includegraphics[width=0.95\linewidth]{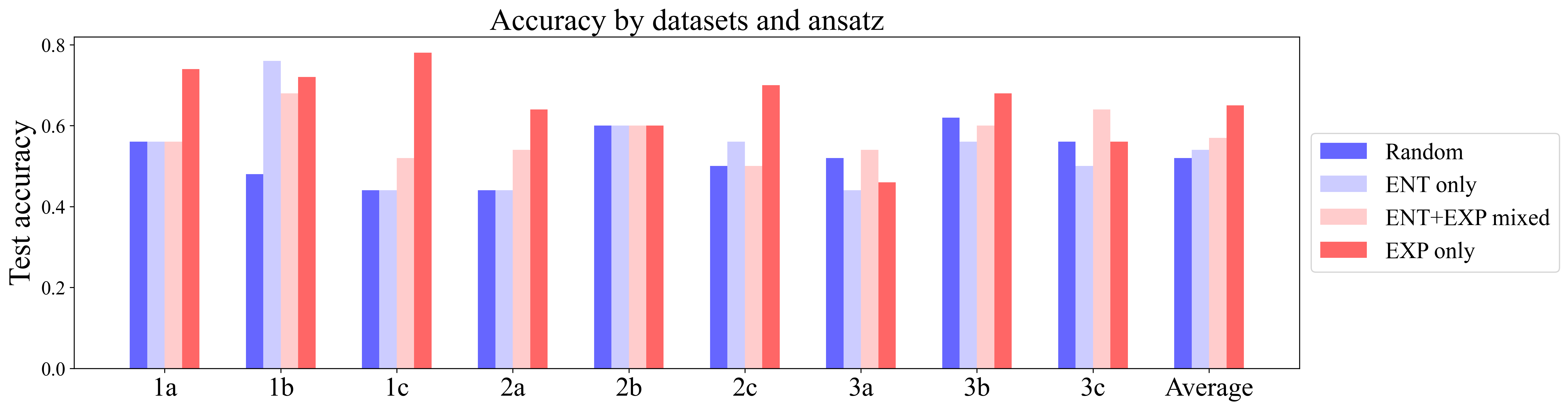}
\caption{Different ways to select the sub-circuits of top performance. ENT stands for sub-circuits with the best entangling capabilities (MW measure); EXP stands for sub-circuits with the best expressibility. We can see from the average accuracy that the expressibility is more correlated to the model accuracy. }
% \vspace{-2mm}
\label{fig_initial_ansatz}
\end{figure*}

\begin{figure*}[h!tb]
\centering
\includegraphics[width=0.95\linewidth]{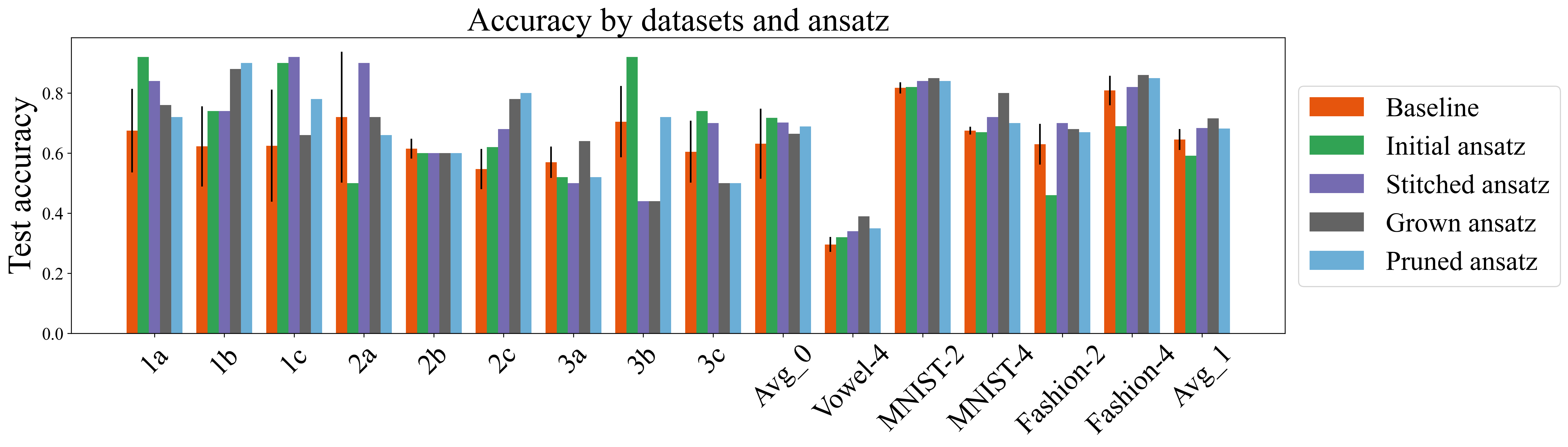}
% \vspace{-7mm}
\caption{The model accuracy at different stages of optimization. Avg\_0 and Avg\_1 stand for averaged results for datasets from ~\cite{hubregtsen2021evaluation} and ~\cite{wang2021quantumnas} respectively.}
% \vspace{-6mm}
\label{fig_main}
\end{figure*}

\subsection{Performance of Combined Sub-circuit}
The topology-aware ansatz is generated by combining sub-circuits with the best criteria. There are several ways to select the sub-circuits with ``good'' properties. We denote the sub-circuits with the best expressibility as ``EXP'' and the sub-circuits with the best entangling capability as ``ENT''. We propose different ways to select the sub-circuits: EXP only, ENT only, ENT+EXP mixed. The baseline is marked as ``random'' and it is a randomly generated ansatz of similar size.
~\cite{hubregtsen2021evaluation} claims that strong correlation exists between classification accuracy and expressibility. At the same time, week correlation exists between the accuracy and the entangling capability. Figure \ref{fig_initial_ansatz} validates this claim. 
The average accuracy for different policies are [0.52, 0.54, 0.57, 0.65]. Our initial ansatz achieves an average accuracy improvement of 13\% compared with random ansatz over nine datasets. It confirms that the selected sub-circuits perform better than random sub-circuits. 

\subsection{Effectiveness of Optimizations}

Figure \ref{fig_main} shows the model accuracy of our ansatz at different optimization stage. We see that the topology-aware ansatz performs better than the manually designed ansatz~\cite{sim2019expressibility}.
% in Figure \ref{fig_base}. 
For datasets \{1a,1b,1c,2a,2b,2c,3a,3b,3c\} adopted from ~\cite{hubregtsen2021evaluation}, we notice a slight drop of accuracy 
after applying the optimization methods.
It indicates that the initial ansatz is already sufficient for such tasks. The initial ansatz show an average 8.6\% advantage over manually-designed ansatz and 13\% advantage over random ansatz in terms of model accuracy. In contrast, for datasets from~\cite{wang2021quantumnas} \{MNIST-2, MNIST-4, FASHION-2, FASHION-4, VOWEL-4\}, we see gradual improvement of model accuracy thanks to the incrementally built ansatz; and the grown ansatz provide an average 7.1\% advantage of accuracy. 
We believe that the difference in problem size causes such phenomenon.

For machine learning tasks of more complexity such as \{MNIST-4, FASHION-4, VOWEL-4\}, the ansatz is larger before accuracy reaches saturation, which is also reflected in Figure \ref{grow_accu}. Since the ansatz generated by our approach is directly compatible with the hardware topology, we expect 
they have smaller size after they are compiled. 
The Table \ref{tab_1} shows that before the compilation process, the size of our ansatz is comparable with the manually-designed ansatz. 
But after the circuits are compiled onto quantum computers, 
our circuits are smaller due to its topology-aware nature. 
Overall, we achieve a depth reduction around 50\% and a CNOT reduction up to 75\%. Considering the fact that CNOT gates generally takes longer than the single-qubit gates on quantum computers. The circuit latency
reduction is in fact more than 50\%. 
% When the circuits are mapped onto the quantum hardware, SWAP gates are needed for the baseline ansatz. Each SWAP gate will be decomposed into 3 CNOT gates. And the controlled-rotation gates used in baseline ansatz will be decomposed into 2 CNOT gates + several single-qubit gates.

% \FloatBarrier

\subsection{Results on Real-world Quantum Computers}
\label{real}

\begin{table*}[]
\caption{Model Accuracy on Different Backends including Non-optimal Initial Layout (NOIL) Results}
\centering
\label{real_res}
\resizebox{0.9\textwidth}{!}{%
\begin{threeparttable}
\begin{tabular}{ccccccccccc}
\hline
 & MNIST-2 & NOIL & MNIST-4 & NOIL & FASHION-2 & NOIL & FASHION-4 & NOIL & Avg & NOIL \\ \hline
noise-free sim & 0.85 & N/A & 0.76 & N/A & 0.84 & N/A & 0.67 & N/A & 0.78 & N/A \\ \hline
ibmq\_quito & 0.86 & 0.87 & 0.71 & 0.63 & 0.84 & 0.80 & 0.69 & 0.65 & 0.78 & 0.74 \\ \hline
ibmq\_lima & 0.86 & 0.71 & 0.76 & 0.67 & 0.83 & 0.83 & 0.68 & 0.67 & 0.78 & 0.72 \\ \hline
ibm\_oslo & 0.87 & 0.58 & 0.75 & 0.31 & 0.84 & 0.77 & 0.65 & 0.50 & 0.78 & 0.54 \\ \hline

\end{tabular}

% \begin{tablenotes}
%   \small
%   \item The accuracy numbers under NOIL columns are obtained with non-optimal initial layout.
% \end{tablenotes}%
\end{threeparttable}

}
\end{table*}

We test the trained ansatz on IBM's cloud quantum computers. Table~\ref{real_res}  
displays different classification accuracy from various backends. Due to the limited size of our designed ansatz circuits, the classification accuracy with real machines are almost identical with the results collected on noise-free simulator. The columns with ``NOIL'' represent the accuracy results with the same ansatz but non-optimal initial layout (NOIL), where extra SWAP gates are needed. We 
insert such mismatch between our ansatz and hardware topology to investigate the affects of extra SWAP gates. For quantum computers (ibmq\_quito) with lower noise, the mismatch introduced slight drops in accuracy. However, it shows sharp decrease of accuracy when tested on quantum devices with higher noise (ibm\_oslo). The difference emphasizes the importance of topology-aware techniques in the NISQ era, where quantum devices are prone to high gate errors. On average, \name shows accuracy advantages by 17\% over the NOIL results.

\subsection{Overhead, Scalability and Barren Plateaus}

The evaluation of sub-circuits is off-line and it only needs to be done once. After the sub-circuits are evaluated, the selection and combination of sub-circuits can be completed with negligible time. 
Therefore, the run-time overhead of our bottom-up approach based on the current quantum computers to create QNN ansatz is very low. As for the training overhead, since we adopt the ``growing '' method, the architecture search for the ansatz will happen during the training process. Under the ansatz of similar size, our overhead is smaller compared with methods where ansatz needs to be re-trained after the structure is determined.
% \violet{DO YOU HAVE ANY DATA?? No, since the major part is sub-circuits evaluation, the steps after the evaluation can be ignored}
On the other side, we expect to see quantum computers of more than one thousand qubits in the next decade\cite{cho2021ibm}. 
While these large computers are likely to have sparser connections,
the graph based on their hardware topology
can always be divided into smaller sub-graphs. 
The major overhead of our bottom-up approach comes from sub-circuits' evaluation, which only scales with the size of sub-circuits. Therefore, the proposed bottom-up approach has high scalability and will work 
well on large quantum computers.

Moreover, recent work~\cite{https://doi.org/10.48550/arxiv.2204.07179} suggests that, ADAPT-VQE, a well-designed ansatz~\cite{grimsley2019adaptive} can naturally preclude the affects from barren plateaus and large numbers of local minima. When additional operators are appended to the ansatz, it is likely
to create a deeper trap to ``burrow'' towards the exact solution.
This explains why our idea of incrementally growing ansatz works, which
is also confirmed in the results.

% \clearpage

\section{Related work}
\label{rel}

% \textcolor{cyan}{
% Q: Greedy incremental construction of ansatz of VQE is used in “An adaptive variational algorithm for exact molecular simulations on a quantum computer”
% }

% \violet{
% Our method is very different from ADAPT-VQE~\cite{grimsley2019adaptive}, where ansatz is also initialized and expanded. We formulate our building blocks based on two criteria, which does not have physical significance.
% }

QML offers various potential applications for small quantum computers~\cite{liang2022variational, sasaki2001quantum,bisio2010optimal,bisio2011quantum,sentis2012quantum,sentis2015quantum,sentis2016quantum,paparo2014quantum,dunjko2015quantum,dunjko2016quantum, cheng2020accqoc}. In recent years, different structures of QNN have been proposed and tested~\cite{liu2021rigorous,peters2021machine,jaderberg2021quantum,wall2021tree}. These works focus on maximizing the accuracy of quantum models for machine learning tasks. At the same time, the quantum compilers 
have made breakthroughs as well~\cite{bertels2020quantum,gokhale2020optimized,li2021co,shi2019optimized,steiger2018projectq}. Better algorithms for qubit mapping and program synthesis have been proposed~\cite{li2019tackling,zulehner2018efficient,niu2020hardware,itoko2020optimization,hillmich2021exploiting,lao2021timing}. Some papers~\cite{wu2020qgo,tan2020optimality} propose to optimize the synthesis process with the information of hardware topology. However, they target  general quantum programs, where the design space is not large. For QNN algorithms, intensive research has been conducted on finding the optimal ansatz~\cite{zhang2021neural,wang2021quantumnas,wang2021roqnn,wang2022chip,nguyen2021quantum,alam2021quantum}.

\section{Conclusion}
\label{con}

In this paper, we propose a bottom-up approach to generate topology-aware ansatz for Variational Quantum Algorithms and we use the important Quantum Neural Networks as the benchmark to evaluate our method.
The effectiveness of this approach is 
due to the flexibility of QNN ansatz.
Because the minimized ansatz does not have physical meaning,
the final states of the ansatz can be from the full Hilbert Space. 
To make the search tractable, we propose to 
first generate hardware compatible 
sub-circuits with ``good'' properties first, then combine the sub-circuits to form the initial ansatz. 
We propose several optimizations to 
compensate for the sparser qubit connection in the initially
generated ansatz and increase accuracy.
With this approach, the search of a huge design space is reduced to 
finding better sub-circuits that can form the high quality ansatz.
We evaluated our approach with 14 data sets used in 
recent works. The results show that 
the ansatz generated by our solution achieves decent model accuracy with ansatz that are 50\% smaller in depth. \name \space on three real machines demonstrates on average 17\% higher accuracy.

\clearpage

\bibliographystyle{plain}
\bibliography{refs}

\end{document}